\documentclass[]{article}
\usepackage[superscript]{cite}
\usepackage{epsfig} 
\def\gsim{ \lower .75ex \hbox{$\sim$} \llap{\raise .27ex \hbox{$>$}} }
\def\lsim{ \lower .75ex\hbox{$\sim$} \llap{\raise .27ex \hbox{$<$}} }

\begin{document}
\begin{center}
{\LARGE \bf  
The power of relativistic jets is larger than the luminosity of their accretion disks
}
\vskip 0.2 cm
{\large
G. Ghisellini$^1$, F. Tavecchio$^1$, L. Maraschi$^2$, A. Celotti$^{1,3,4}$, T. Sbarrato$^{1,5,6}$\\
}
\end{center}
\vskip 0.2 cm
\noindent
1. INAF -- Osservatorio Astronomico di Brera, via E. Bianchi 46, I--23807, Merate, Italy \\
2. INAF -- Osservatorio Astronomico di Brera, via E. Brera 28, I--20121, Milano, Italy \\
3. SISSA, Via Bonomea 265, I--34135, Trieste, Italy \\
4. INFN--Sezione di Trieste, via Valerio 2, I-34127 Trieste, Italy \\
5. Univ. dell'Insubria, Dipartimento di Fisica e Matematica, Via Valleggio 11, I--22100 Como, Italy \\
6. ESO--European Southern Observ., Karl--Schwarzschild-Strasse 2, 8578 Garching bei M\"unchen, Germany\\

\large

{\bf
Theoretical models for the production of relativistic jets from active galactic nuclei 
predict that jet power arises from the spin and mass of the central black hole, 
as well as the magnetic field near the event horizon{\cite{blandford78}}. 
The physical mechanism mechanism underlying the contribution from the magnetic field 
is the torque exerted on the rotating black hole by the 
field amplified by the accreting material.
If the squared magnetic field is proportional to the accretion rate,
then there will be a correlation between jet power and accretion luminosity. 
There is evidence for such a 
correlation{\cite{rs91,celotti93,celotti97,maraschi03,punsly06,celotti08,gg10general}},
but inadequate knowledge of the accretion luminosity of the limited
and inhomogeneous used samples prevented a firm conclusion.
Here we report an analysis 
of archival observations of a sample of blazars (quasars whose jets point towards Earth) 
that overcomes previous limitations.
We find a clear correlation between jet power as measured through the $\gamma$--ray luminosity,
and accretion luminosity as measured by the broad emission lines, with the jet power
dominating over the disk luminosity, in agreement with numerical simulations{\cite{tchekhovskoy11}}.
This implies that the magnetic field threading the black hole horizon reaches the maximum value
sustainable by the accreting matter{\cite{zamaninasab14}}.
}

\vskip 0.5 cm

The jet power is predicted\cite{blandford78} to depend on $(a M B)^2$, where $a$ and $M$ are
respectively the spin and mass of the black hole and $B$ is the magnetic field at its horizon. 
Seed magnetic fields are amplified by the accretion disk
up to equipartition with the mass energy density $\sim \rho c^2$ of the matter accreting 
at the rate $\dot M$. 
A greater $\dot M$ implies a larger $\rho$, that can sustain a larger magnetic field, 
which in turn can tap a larger amount of the black hole rotational energy.
The magnetic field is thus a {\it catalyst} for the process. 
Increasing the spin of the black hole shrinks the innermost stable orbit, 
increasing the accretion efficiency $\eta$ (defined by $\eta=L_{\rm disk}/\dot M c^2$) 
($L_{\rm disk}$ accretion disk luminosity) to a maximum value{\cite{thorne74}} $\eta=0.3$. 

We use a well designed sample of blazars that have been
detected in the $\gamma$--ray band by the {\it Fermi} Large Area Telescope (LAT) 
that have been spectroscopically observed in the optical{\cite{shaw12,shaw13}}  (see Methods). 
They have been classified as BL Lac objects or Flat Spectrum Radio Quasars 
(FSRQs) according if the rest frame equivalent width of their broad emission lines was 
greater (FSRQ) or smaller (BL Lacs) than 5 \AA\ (rest frame).
The sample contains 229 FSRQs, and 475 BL Lacs. 
Of the latter, 209 have a spectroscopically measured redshift. 
We considered all FSRQs with enough multi--wavelength data to have 
a Spectral Energy Distribution (SED) that allows to establish the 
bolometric luminosity.
The considered FSRQs amount to 191 objects.
Instead, for BL Lacs, we consider only the 26 sources with detected 
broad emission lines.
This makes them the low disk luminosity tail of of the full blazar sample.
This choice is dictated by our will to measure the accretion luminosity,
together with the jet power.
Through the visible broad emission lines we reconstruct,
through a template{\cite{francis91,vanderberk01}}, 
the luminosity of the entire broad line region.
The latter is a proxy of the accretion disk luminosity, 
$L_{\rm BLR} = \phi L_{\rm disk}$, with\cite{calderone13} $\phi\sim 0.1$.
This disk luminosity $L_{\rm disk}$ is then directly given by the observed 
broad emission lines, avoiding the contamination by the non--thermal continuum.
Uncertainties are admittedly large (factor $\sim$2) for specific
sources, but the averages should be representative of the true values.

To model the non--thermal jet emission, we applied to all objects a simple one--zone leptonic 
model\cite{gg09} (see Methods), from which we derive the physical parameters of the jet. 
The only parameter of interest here, however, is the bulk Lorentz factor $\Gamma$
of the outflowing plasma, found in the range 10--15 (see Methods and Extended Data Fig. 2).
This range is
similar to that obtained via measurements of the superluminal motion of the 
radio components, which however occurs on larger distances from the black hole.
The bulk Lorentz factor is thus only weakly model--dependent.
Having the bolometric jet luminosity $L^{\rm bol}_{\rm jet}$ and $\Gamma$,
the power that the jet spent in producing the non thermal radiation 
is{\cite{gg10}}:   
\begin{equation}
P_{\rm rad} \, = \, 2\,f {L^{\rm bol}_{\rm jet} \over \Gamma^2}
\label{pr}
\end{equation}
where the factor 2 accounts for two jets and $f$, of order unity,
is discussed in the Methods section.
If this were the entire power of the jet, it would be entirely spent
to produce the observed radiation.
The jet would stop, and could not produce the radio lobes and/or
the extended radio emission we see from these objects.
It is then a strict {\it lower limit} of the jet power.

\begin{figure}
\includegraphics[height=0.75\textheight]{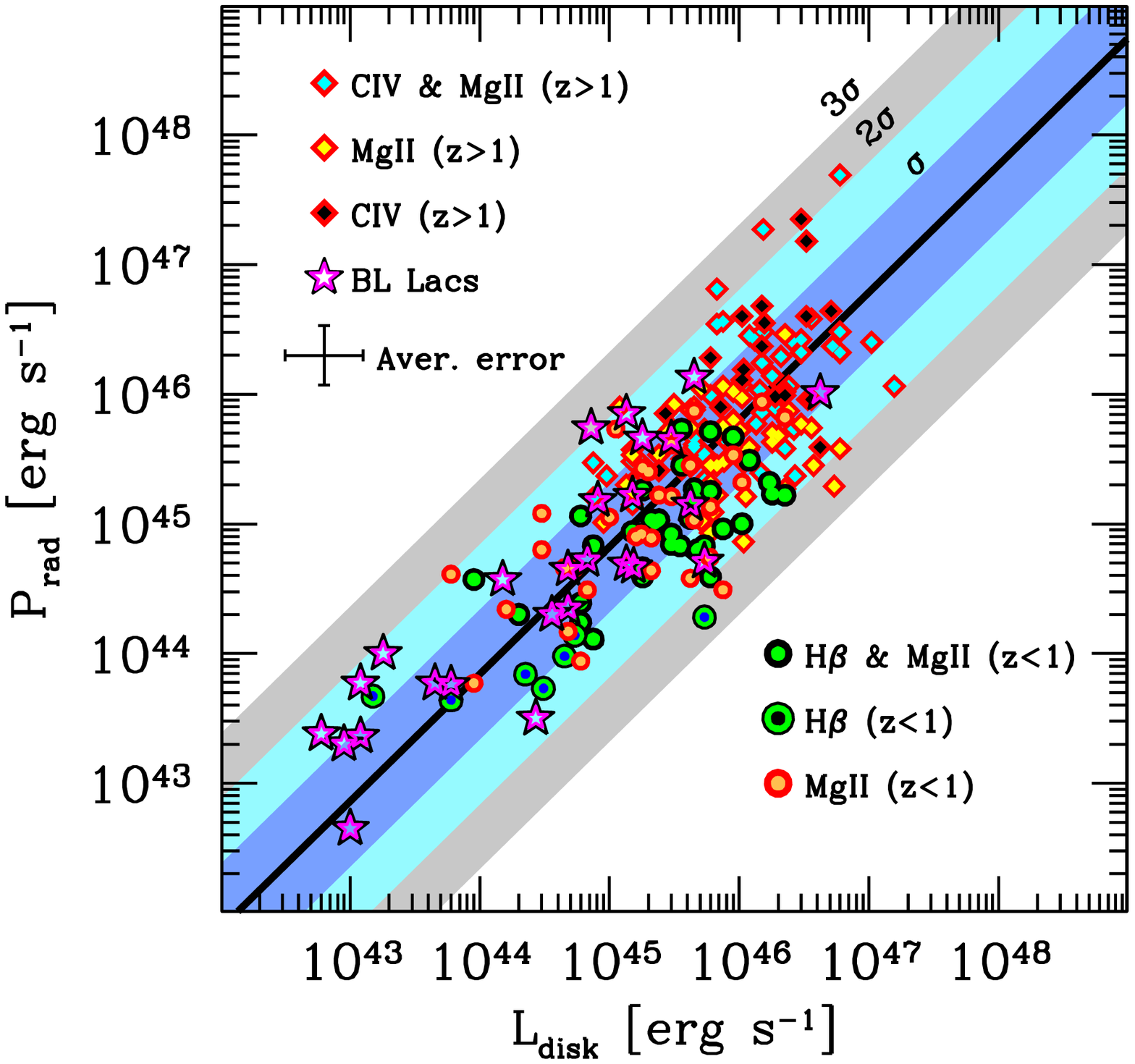}\\
\caption{\large
{\bf 
Radiative jet power vs disk luminosity. 
}
The radiative jet power
versus the disk luminosity, calculated as ten times the luminosity of the Broad Line Region.
Different symbols correspond to the different emission lines
used to estimate the disk luminosity, as labelled. 
All objects have been detected by {\it Fermi}/LAT and have been spectroscopically 
observed in the optical{\cite{shaw12,shaw13}}.
Shaded colored areas correspond to 1, 2 and 3 $\sigma$ (vertical) dispersion. $\sigma=0.5$ dex.
The black line is the best least square fit 
($\log P_{\rm rad}=0.98 \log L_{\rm disk}+0.639$). 
The average error bar corresponds to an uncertainty of a factor 2 in $L_{\rm disk}$\cite{calderone13}
and 1.7 in $P_{\rm rad}$ (corresponding to the uncertainty in $\Gamma^2$).
}
\label{prld}
\end{figure}

Fig. \ref{prld} shows $P_{\rm rad}$ as a function of $L_{\rm disk}$
for all the 217 considered blazars. 
There is a robust correlation between the two:
$\log P_{\rm rad}=0.98\log L_{\rm disk} + 0.639$ (with a probability $P<10^{-8}$
to be random, even taking into account the common redshift dependence).
We thus find a linear correlation between the minimum jet power and the accretion
luminosity, as expected.
Moreover, the two are of same order.
This holds also for the considered BL Lacs that do show broad emission lines.
The dispersion along the fitting line is $\sigma=0.5$ dex.
An important contribution to this dispersion comes from the large amplitude
variability of the non--thermal flux displayed by all blazars, especially 
in the $\gamma$--ray band, where the bolometric jet luminosity peaks.
This is true even if we considered the LAT luminosity averaged over two years{\cite{nolan12}},
as shown by the comparison between LAT and older Energetic Gamma Ray Experiment Telescope 
(EGRET, onboard the {\it Gamma Ray Compton Observatory}) 
results: about 20\% of the EGRET detected blazars
are not detected by LAT{\cite{ghirlanda11}}, even though the latter has a 20--fold better sensitivity.

The power in radiation $P_{\rm rad}$ is believed to be about 10\% of the jet power $P_{\rm jet}$
and, remarkably, this holds true both for Active Galactic Nuclei (AGN) and Gamma--Ray Bursts{\cite{nemmen12}}.
We confirm this result, if there is one proton per emitting lepton (see Methods and
Extended data Figure 1).
This limits the importance of electron--positron pairs,
that would reduce the total jet power. 
In addition, pairs cannot largely outnumber protons, 
since otherwise the Compton rocket effect would stop the jet{\cite{gg10}} (see Methods). 

An inevitable consequence of $P_{\rm jet} \sim 10 P_{\rm rad}$ is that 
the jet power is larger than the disk luminosity.
Therefore the process that launches and accelerates jets must be
extremely efficient, and might be the {\it most} efficient way of transporting
energy from the vicinity of the black hole to infinity.

\begin{figure}
\includegraphics[height=0.75\textheight]{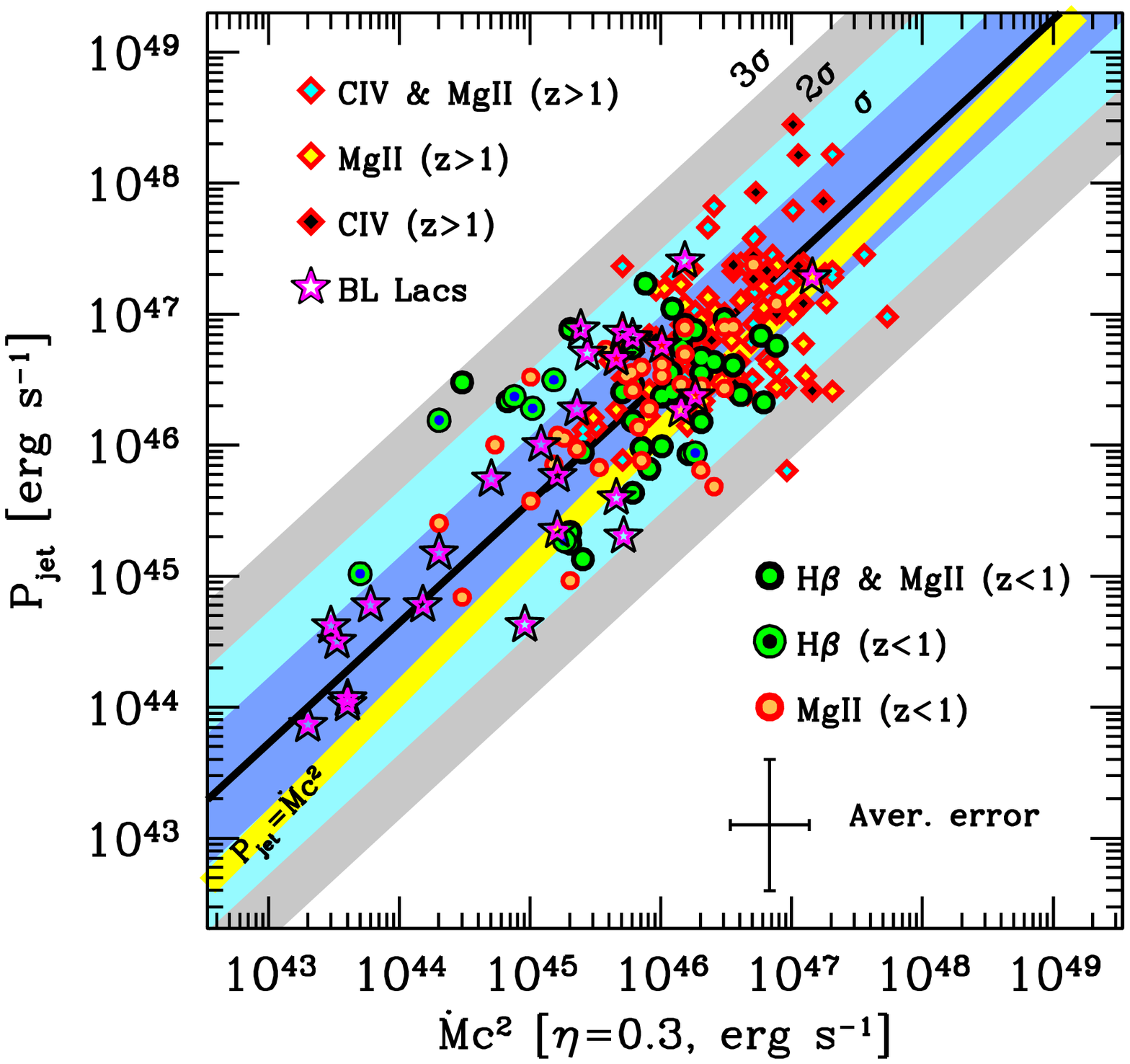}
\caption{\large
{\bf
Jet power vs accretion power. 
}
The total jet power estimated through a simple one--zone leptonic model{\cite{gg09}}, 
assuming one cold proton per emitting electron, 
vs $\dot M c^2$ calculated assuming an efficiency $\eta=0.3$, 
appropriate for a maximally rotating Kerr Black hole.
Different symbols correspond to the different emission lines
used to estimate the disk luminosity, as in Fig. \ref{prld}. 
Shaded colored areas correspond to 1, 2 and 3 $\sigma$ (vertical) dispersion. $\sigma=0.5$ dex.
The black line is the best least square fit 
[$\log P_{\rm jet}=0.92 \log (\dot M c^2)+4.09$].
The yellow stripe is the equality line.
The average error bar is indicated ($\dot Mc^2$ has the same average uncertainty of $L_{\rm disk}$;
the average uncertainty $P_{\rm jet}$ is a factor 3).
}
\label{pjdotm}
\vskip 3 cm
\end{figure}

Assuming $\eta=0.3$, appropriate for fastly rotating black holes,
we have $\dot M c^2 =L_{\rm disk}/\eta$. 
Fig. \ref{pjdotm} shows the jet power $P_{\rm jet}$ vs $\dot M c^2$ for
all our sources.
The white stripe indicates equality. 
The black line is the best fit correlation 
[$\log P_{\rm jet}=0.92 \log (\dot M c^2)+4.09$]
and lies always above the equality line. 
This  finding is fully consistent with recent general relativistic magnetohydrodynamic
numerical simulations{\cite{tchekhovskoy11}},
in which the average outflowing power in jets and winds reaches 140\% of $\dot M c^2$
for dimensionless spin values $a=$0.99.
The presence of the jet implies that the gravitational potential energy of
the falling matter can not only be transformed into heat and radiation, but can
also amplify the magnetic field, allowing the field to access 
the large store of black hole rotational energy and 
transform part of it into jet mechanical power.
This jet power is  somewhat larger than 
the entire gravitational power $\dot M c^2$ of the accreting matter.
This is not a coincidence, but the result of the catalyst role of the magnetic 
field amplified by the disk.
When the magnetic energy density exceeds the energy density $\sim \rho c^2$ of
the accreting matter in the vicinity of the last stable orbit, the
accretion is halted and the magnetic energy decreases, as shown
by numerical simulations{\cite{tchekhovskoy11,mad}},
and confirmed by recent observational evidence{\cite{zamaninasab14}}.

The mass of the black holes of the FSRQs in our sample 
has been calculated{\cite{shaw12}} assuming that the size of the broad line 
region scales with the square root of the ionizing disk luminosity
as indicated by reverberation mapping{\cite{peterson00,mclure04}}, and by  
assuming that the clouds producing the broad emission lines are virialized.
The uncertainties associated to this method are large
(dispersion of $\sigma$=0.5 dex for the black hole mass values\cite{vestergaard06}),
but if there is no systematic error (see Methods) the average Eddington ratio for FSRQs
is reliable: $\langle L_{\rm disk}/L_{\rm Edd}\rangle=0.1$ (Extended Data Figure 2).
This implies that all FSRQ should have standard, geometrically thin, optically thick, 
accretion disks{\cite{shakura73}}.  
Therefore the more powerful jets (the ones associated to FSRQs) 
{\it can} be produced by standard disks with presumably no central funnel, 
contrary to some expectations{\cite{livio99,meier02}}.

A related issue is the possible change of accretion regime at
low accretion rate (in Eddington units), or, equivalently,
when $L_{\rm disk}\lsim 10^{-2}L_{\rm Edd}$.
In this case the disk is expected to become radiatively inefficient,
hotter and geometrically thick.
How the jet responds to such changes is still an open issue.
An extension of our study to lower luminosities could provide some hints.
Another open issue is how the jet power depends on the black hole spin{\cite{tchekhovskoy12}}.
Our source sample by construction consists of luminous $\gamma$--ray sources that 
presumably have the most powerful jets, and thus have the most rapidly spinning holes.
It will be interesting to explore less luminous jetted sources, to get hints
on the possible dependencies of the jet power on the black hole spin and the possible 
existence of a minimum spin value for the very existence of the jet.
In turn, this should shed light on the long standing problem of the
radio--loud/radio--quiet quasar dichotomy{\cite{sikora07}}.

\vskip 2 cm
\noindent
{\bf Acknowledgements}
FT and LM thank a PRIN--INAF 2011 grant for partial funding.

\vskip 0.5 cm
\noindent
{\bf Author Contributions}
GG wrote the manuscript and fitted all blazars presented. 
FT, LM, AC and TS contributed to the discussion for the implications
of our results.

\vskip 0.5 cm
\noindent
{\bf Author information}
Reprints and permissions information is available  at \\
www.nature.com/reprints.
The authors declare no competing financial interests. \\
Correspondence and requests for materials should be addressed to G.G. \\
(gabriele.ghisellini@brera.inaf.it).

\vskip 3 cm

\noindent
{\large {\bf METHODS}}
\vskip 0.3 cm
\noindent
{\bf The sample.} 
Our sample is composed by 229 blazars detected by {\it Fermi}/LAT{\cite{abdo10,ackermann11}}
for which broad emission lines have been measured{\cite{shaw12,shaw13,shen11}}.
This sample does not include several bright and famous blazars
with historical spectroscopic classifications in the literature.
Of these, we have studied the 217 objects with enough multiwavelength 
information necessary to apply our model.
Within the sample, we have 26 BL Lac objects
with measured broad emission lines{\cite{shaw13}}.
Therefore our ``BL Lac" objects, even if fulfilling the classical 
definition of BL Lacs (emission lines of equivalent width smaller than 5 \AA), are
not lineless, and can be considered as the low disk luminosity tail of the blazar sample.

This is the largest ever sample of $\gamma$--ray detected sources with measured
broad emission lines. 
From these lines we can estimate the luminosity $L_{\rm BLR}$ of the entire broad line region (BLR), 
using standard templates{\cite{francis91,vanderberk01}}.
Then we calculate the disk luminosity by assuming that $L_{\rm disk} = 10 L_{\rm BLR}$,
with an average uncertainty\cite{calderone13} of a factor 2.
Since the lines are isotropically emitted, the estimate of $L_{\rm disk}$
does not depend on the viewing angle. 
Moreover, $L_{\rm disk}$ estimated from $L_{\rm BLR}$ does not depend
on any contamination of non--thermal components. 
In a few cases (20 out of 217, Extended Data Table 1) 
the resulting $L_{\rm disk}$ contrasts by a factor between 2 and 5
with the value of $L_{\rm disk}$ given by fitting a standard accretion 
disk, that better fits the optical--UV data.
In these cases we have chosen the value of $L_{\rm disk}$ given by the disk fitting.
From the knowledge of the SED, often dominated by the $\gamma$--ray luminosity,
we can estimate the bolometric jet luminosity in a reliable way.
The knowledge of the disk luminosity and the black hole mass greatly helps to
fix two important parameters for the theoretical modelling, helping to find
a unique solution for the considered emitting region of the jet.
Although the mass estimate can be affected by a rather large statistical error, 
there should be no relevant systematic error, since a completely independent
method\cite{massefit1,massefit2} led to consistent values.

\vskip 0.5 cm
\noindent
{\bf 
The bolometric jet luminosity.
}
The spectral energy distribution (SED) of blazars is characterized
by two broad humps [in the $\nu L(\nu)$ vs $\nu$ representation, where $\nu$ is
the emission frequency],
peaking in the mm--UV and in the MeV--GeV bands.
The high energy bump is often the dominant component, except for 
low power (and lineless) BL Lacs{\cite{fossati98,donato01}}, in which 
the synchrotron luminosity is more important.
This is the reason to select {\it Fermi}/LAT detected blazars:
for these objects we can asses the jet bolometric luminosity
with high confidence.
However, the amplitude of variability, especially in the $\gamma$--ray
band, can be larger than 2 orders of magnitude, and even larger if one
includes exceptionally high states, as the ones experienced by 
3C 454.3{\cite{bonnoli11}}.
Taking the mean luminosity over a period  of two years 
averages out short term variability, but not the secular ($>$10 yr timescale) one.
The $\gamma$--ray luminosity measured in {\it detected} sources could 
not represent the average status of the source, but only its high state.
On the other hand, 
the $\gamma$--ray luminosity correlates with the radio one{\cite{ghirlanda11}},
and a Gaussian distribution of long term $\gamma$--ray 
variability with $\sigma=0.5$ dex (i.e. a factor 3), coupled with 
the {\it Fermi}/LAT sensitivity, can fully
explain what is observed, including the fact that several strong radio sources are
yet undetected in the $\gamma$--rays.
As a result, the observed correlations in Figg. \ref{prld} and \ref{pjdotm}
could represent jet--active states, rather then the average state,
that could correspond to a jet power $P_{\rm jet}$ up to a factor $\sim$3 smaller.
However, the radio--$\gamma$ correlation gives us confidence that $P_{\rm jet}$
would still be correlated with $L_{\rm disk}$ and $\dot M$.
The fact that the process of forming and launching relativistic jets
is more powerful than accretion only in jet--active states does not affect 
the conclusion that this process is indeed more powerful than accretion, even 
if it does not always works at its maximum pace.

\vskip 0.5 cm
\noindent
{\bf 
The model used.
}
We summarize here the main features of the used model{\cite{ghisellini09}}.
It assumes that the jet region emitting most of the 
non--thermal luminosity is at a distance $R_{\rm diss}$
from the black hole. For this reason this class of models is called ``one--zone",
and they are justified because often (although not always) we 
see coordinated variability at different frequency bands.
The jet is assumed conical with semi--aperture angle $\psi$.
We assume $\psi=0.1${\cite{nalewajko14}}, but the exact value is not critical
for our results.
The emitting region is assumed spherical, with radius $R=\psi R_{\rm diss}$,
embedded in a homogeneous but tangled magnetic field $B$.
The emitting particles are leptons (requiring less power than the less common alternative
of emitting hadrons\cite{bottcher13}).
The main feature of the model is that it accounts for the 
radiation fields produced by the emission disk, the broad line region, 
the dusty torus surrounding the disk and re--emitting part of the
intercepted radiation in the infrared.
The distance of the BLR is assumed to be a function of the 
disk luminosity, as indicated by recent observations (through the so--called reverberation mapping
technique\cite{bentz06}):
$R_{\rm BLR} = 10^{17} L_{\rm d, 45}^{1/2}$ cm.
We assume that  also the typical size of the molecular torus scales similarly:
$R_{\rm torus}= 2\times 10^{18} L_{\rm d, 45}^{1/2}$ cm (in agreement with very recent
reverberation results\cite{koshida14}).
Here $L_{\rm d,45}$ is the disk luminosity in units of $10^{45}$ erg s$^{-1}$.
As a consequence, inside the BLR, the radiative energy density corresponding to broad lines is
constant: $U_{\rm BLR} \sim 0.1 L_{\rm disk} / [ 4\pi R^2_{\rm BLR}c] = 1/(12 \pi)$ erg cm$^{-3}$.
A similarly expression holds for the energy density of the IR photons of the torus.

The particle distribution responsible for the produced radiation is derived
through a continuity equation assuming continuous injection of relativistic leptons at $R_{\rm diss}$,
their radiative cooling, the possible production of electron--positron pairs
through photon--photon collisions and their corresponding radiation.
The energy distribution of the injected particles is a broken power--law,
flat at low energies and steepening above some break energy $\gamma_{\rm b}$.
Since the considered emitting region is always compact, its self--absorption frequency
is always large, and the model cannot account for the radio flux at observed frequencies
smaller than a few hundreds GHz. 
These are produced by the superposition of several, larger, components. 
The emission produced by the accretion disk is assumed to be a multicolor blackbody,
with a temperature distribution dictated by the balance of heat production and radiative 
dissipation{\cite{shakura73}}.
The corresponding values of $L_{\rm disk}$ found through disk fitting 
are listed in the Extended Data Table 1.

Although the model returns several physical parameters, we concentrate on the
ones of interest here: the bulk Lorentz factor $\Gamma$ and Doppler factor $\delta$, 
the location of the emitting region and the jet power.

\vskip 0.5 cm
\noindent
{\bf 
The bulk Lorentz factor $\Gamma$ and Doppler factor $\delta$.
}
The bulk Lorentz factor, coupled with the viewing angle $\theta_{\rm v}$,
determines the Doppler factor $\delta\equiv 1/[\Gamma(1-\beta\cos\theta_{\rm v})]$.
For blazars, we have $\sin\theta_{\rm v}\sim 1/\Gamma$ and thus $\delta\sim\Gamma$.
There are several observables affected by $\Gamma$:

i) The observed $\nu F(\nu)$ flux is amplified by a factor $\delta^4$ 
with respect to the comoving value for the synchrotron (Syn)
and the self Compton (SSC) emission, and more for the flux produced through scattering 
with photons produced external to the jet\cite{dermer95} (this is because, in the
comoving frame, the external seed photons are not isotropic, but
coming from the forward direction).

ii) As long as the emitting region is inside the BLR, the corresponding energy density
is amplified by a factor $\sim \Gamma^2$ (independently of $\theta_{\rm v}$).
Similarly for the IR emission coming from the torus.
If $R_{\rm diss}<R_{\rm BLR}$, the external Compton (EC) process  most likely
dominates over the SSC process, and the Compton to synchrotron luminosity ratio
(equal to the radiative to magnetic energy density ratio in the comoving frame:
$U^\prime_{\rm BLR} / U^\prime_B$) becomes proportional to $(\Gamma/B)^2$.
The same occurs for the IR radiation reprocessed by the torus as long as
$R_{\rm diss}<R_{\rm torus}$.

iii) The Doppler boosting 
regulates the importance of the SSC component. 
In brief: the larger the $\delta$ factor, the smaller the synchrotron radiation energy 
density in the comoving frame, and therefore the smaller the SSC component.

iv) The Doppler factor blueshifts the observed peak frequencies.

In conclusion, there are several observables that depend on combinations of
parameters that include the $\Gamma$ and the $\delta$ factors. 
Finding the best representation of the data thus implies to find a preferred value
for these parameters.

By modelling the 217 blazars of our sample we find
a rather narrow distribution of the bulk Lorentz factors, peaking at $\Gamma\sim 13$
(Extended Data Figure 2).
A Gaussian fit returns a dispersion of $\sigma=1.4$.
The (few) studied BL Lac objects do not show any difference from FSRQs.

\vskip 0.5 cm
\noindent
{\bf 
The jet power. 
} 
The power carried by the jet is in different forms, calculated as:
\begin{equation}
P_{\rm i} \, =\,2\, \pi R^2\Gamma^2 c\, U^\prime_{\rm i}
\end{equation}
where the factor 2 accounts for having two jets
and $U^\prime_{\rm i}$ is the comoving energy density of protons (i=p), relativistic electrons (i=e),
magnetic field (i=$B$) and of the produced radiation (i=rad).
The radiative power $P_{\rm rad}$ can also be found through{\cite{gg10}}:
\begin{eqnarray}
P_{\rm rad} \, &=& \, 2 \times \,{4\over 3} L^{\rm obs}_{\rm bol, jet} 
{\Gamma^2\over \delta^4} , \,\, \quad {\rm EC}
\nonumber \\
P_{\rm rad} \, &=& \, 2 \times \,{16\over 5}  L^{\rm obs}_{\rm bol, jet} 
{\Gamma^4\over \delta^6} ,
\quad {\rm Syn\,\, and\,\, SSC} 
\end{eqnarray}
This is the way $P_{\rm rad}$ has been calculated: Eq. \ref{pr} is the 
approximation for $\delta\sim \Gamma$, and where $f$ corresponds to the 
numerical factor (4/3) or (16/5).

The main logic is: by applying a radiative model, we derive how much magnetic field and 
how many emitting leptons are required to account for the observed luminosity, and also 
the size and the bulk Lorentz factor of the emitting region.
We then assume that {\it all} leptons present in the jet participate to the emission,
and that for each lepton there is a proton. 
We assume them cold, even if shock acceleration and/or
magnetic reconnection would give at least equal energy to the leptons and to the protons.
This simplification is reasonable as long as the average electron energy remains smaller 
than the proton rest--mass.

The power $P_{\rm rad}$ is a {\it lower limit} to the power
because if the total jet power were $P_{\rm rad}$, then it would convert
all its bulk kinetic energy to produce the radiation we see, and it would stop,
well before we see it still moving, i.e. with Very Large Baseline Interferometry
(VLBI) observations, that samples a region pc away from the black hole.
The distributions of the different forms of the jet power are shown 
in the Extended Data Fig. 3, where they are compared with the
distribution of $L_{\rm disk}$.
To account for $P_{\rm rad}$, the Poynting flux and $P_{\rm e}$ are not sufficient. 
We do need another form of power.
The simple solution is to assume that the jet carries enough protons to have 
$P_{\rm p}>P_{\rm rad}$.
This is strengthened by the fact that if the jet were made up of pairs only, it would suffer 
a strong deceleration due to the Compton rocket effect when crossing the broad line region, and
it would stop.
In fact, in the comoving frame of the jet, the external photons are seen coming from the
forward direction. Even if the leptons are distributed isotropically,
head--on scatterings along the forward direction of the jet axis
would be more frequent and energetic than tail-on scatterings. 
The produced radiation, in the jet comoving frame,
has a forward momentum, compensated by a recoil of the jet emitting region.
With no protons, the jet strongly decelerates.
Not to have a strong (i.e. halving $\Gamma$) deceleration{\cite{gg10}},
the number of pairs should not exceed $\sim$10--20 per proton,
in agreement with estimates made with independent 
methods{{\cite{sikora00,gg10,celotti08}}.

\vskip 0.5 cm
\noindent
{\bf 
Jet power and location of the emitting region. 
} 
The location of the emitting region could impact on the required jet power.
The emitting region is estimated to be at distances $R_{\rm diss}<R_{\rm BLR}$
(85\% of the sources) or at $R_{\rm BLR}<R_{\rm diss}<R_{\rm torus}$ (15\% of the sources).
This is dictated by the SED properties
(i.e. if the Compton peak is at $\sim$MeV energies a smaller frequency
seed external field is preferred).
Locating the source much further out, where there are no important sources
of external photons, would increase the jet power requirements, as shown below.
We have the following two possibilities:

1) the SED could result from Synchrotron+SSC. In this case the parameters can
be found univocally\cite{tavecchio98}}.
The synchrotron $\nu_{\rm S}$ and Compton $\nu_{\rm C}$ peak frequencies ratio gives $\gamma_{\rm peak}^2$.
The Compton dominance (Compton to synchrotron luminosity ratio $L_{\rm C}/L_{\rm S}$) 
and the definition of $\nu_{\rm S}=3.6\times 10^6 \gamma^2_{\rm peak} \delta/(1+z)$ give:
\begin{equation}
   B\delta^2 = { L_{\rm syn} \over R} \left[ {2\over c L_{\rm c} } \right]^{1/2}; \qquad 
   B \delta= { \nu_{\rm S}^2 \over 3.6\times 10^6 \nu_{\rm C} (1+z)}
\end{equation}
Solving for $B$ and $\delta$, (setting $R= c t_{\rm var} \delta/(1+z)]$ and
inserting typical values of the observables (i.e. 
$t_{\rm var}\sim$1 day, 
$\nu_S\sim 10^{13}$ Hz, 
$\nu_{\rm C}\sim 10^{22}$ Hz, 
$L_{\rm S}\sim 10^{46}$ erg s$^{-1}$ and
$L_{\rm C}\sim 10^{47}$ erg s$^{-1}$)
one finds small $B$ ($\lsim10^{-4}$ Gauss) and large $\delta$ ($\gsim$100). 
A large $\delta$ in turn requires very small viewing angles
($<1^\circ$, causing problems with determining the number of the sources belonging 
to the parent population) and large $\Gamma$.
As a consequence, the energy densities inside the source are very small,
making the cooling time very long. Invoking the second order SSC does not help,
since all comoving radiation energy densities are small, because $\delta$ is large.
The pure SSC process, applied to the sources in our sample, is thus very inefficient.
This implies that more emitting electrons are needed to produce the observed flux, even accounting
for the larger beaming.
$P_{\rm rad}$ is small (because it is $\propto \Gamma^{-2}$), but $P_{\rm e}$ is increased.
The source is away from equipartition, and its total minimum power
is greater than it would be if the source were in a dense external photon 
environment{\cite{ghisellini01}}.
We have directly experimented this by applying the pure SSC model to some
sources. In the case of 0325+2224, at $z$=2.066, we derive:
Log $P_{\rm rad}$=44.9, Log $P_{\rm p}$=48.6, Log $P_{\rm e}$=47.3 and Log $P_{\rm B}$=42.4, 
to be compared with the values in Extended Data Table 1.
%
Far from sources of external photons (i.e. pure SSC) the required total jet power 
{\it increases}.  
Having two emitting regions (one for the synchrotron, another for the inverse Compton components)
does not help, beacuse the component emitting the $\gamma$--rays 
must produce less synchrotron radiation than we see,
requiring an even smaller magnetic field: the radiative
cooling is even less, and the entire process is even less efficient.
Since the probability that a single emitting region is aligned within $1^\circ$
to the observer (as required by the large delta) is very small, 
there must be several of these small regions pointing in different directions. 
The power taht we calculate on the basis of observations refers to only one of these regions.  
The total jet power is bound to be much more.

2) The SED is made by a Spine/Layer structure, i.e. a slow layer
surrounding a fast spine.  The layer emits, and its photons are
scattered by the spine, enhancing its Compton flux with respect to
the pure SSC case. The radiative cooling is then more efficient. 
This model is very similar to the one we used, with one difference:
using the external photons made by the BLR and the torus is ``for free",
while using the photons made by the layer implies that the jet puts some
energy and power also in the layer, besides in the spine. 
This model therefore inevitably implies a more powerful jet.

\vskip 0.5 cm
\noindent
{\bf 
Low energy electrons and jet power. 
} 
The energy distribution of injected electrons has a flat slope ($\propto \gamma^{-{s_1}}$)
at low energies, with $-1<s_1<1$.
We then calculate the particle distribution $N(\gamma)$ at the time $R_{\rm blob}/c$,
and the cooling energy $\gamma_{\rm cool}$ at this time.
Electrons of energy $\gamma<\gamma_{\rm cool}$ retain the injected slope.  
Due to the flat $s_1$, the number of electrons between $\gamma\sim 1$ and $\gamma_{\rm cool}$  
is small compared to the number of electrons above $\gamma_{\rm cool}$.
In the EC scenario, low energy electrons are responsible for the X--ray spectrum,
so the value of $\gamma_{\rm cool}$ is constrained by the data.
The total number of emitting electron is well constrained.

\vskip 0.5 cm
\noindent
{\bf 
Electron positron pairs and jet power. 
} 
If electron positron pairs are present, the number of protons is reduced,
with a corresponding reduction of the jet power, up to a factor 10 
(not to suffer a too strong Compton rocket effect).
On the other hand, producing the required number of pairs is problematic.
In fact they {\it cannot} be produced at $R_{\rm diss}$. 
They would be relativistic from the start 
and emit  X--rays, filling the``valley" in the X--ray part
of the SED{\cite{ghisellini96}}.
They {\it cannot} be produced by the accretion disk, which is too cold.
The only possible source is the initial, accelerating
part of the jet, whose observed radiation is overwhelmed by the much more 
beamed flux produced at $R_{\rm diss}$.  
This possibility requires a self absorbed synchrotron flux and quasi--thermal 
Comptonization making the spectrum to peak exactly at 1 MeV{\cite{ghisellini12}}.
A peak at higher energies implies pairs born relativistic and fewer in number; a peak
below the pair production threshold $m_{\rm e}c^2$ implies too few produced pairs.
We find this scenario rather ad hoc and very unlikely to occur in all sources.

Consider also that:

i) If $P_{\rm jet}$ is lowered by a factor 10 then
almost half of the sources in our sample would have $P_{\rm jet} < P_{\rm rad}$ (Fig. 1):
this implies that the jet stops at $R_{\rm diss}$
(and no radio halo, no superluminal motion would be possible);
ii) It is found that $P_{\rm jet}\sim 10 P_{\rm rad}$ for blazars and for Gamma Ray Burst, 
using arguments completely different 
from ours{\cite{nemmen12}}.

 \vskip 1 cm
\noindent
{\large \bf EXTENDED DATA}

\setcounter{figure}{0}
\begin{figure}
\includegraphics[height=0.7\textheight]{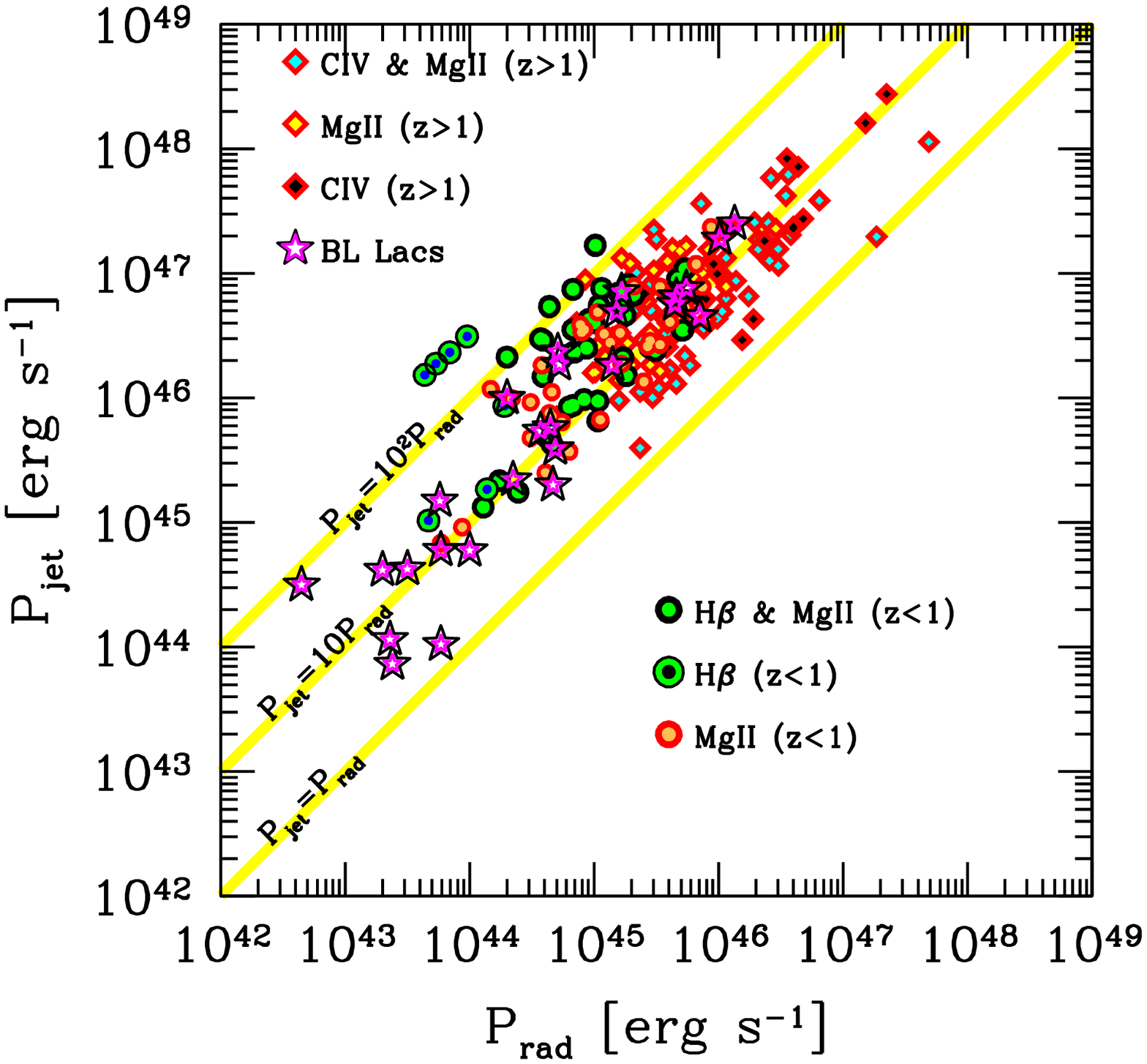} 
\caption{\large
{\bf 
Extended Data Figure 1: Jet power vs radiative jet power. 
}
We compare the total jet power and the radiative one for the blazars
in our sample. The yellow lines, as labelled, correspond to 
equality and to $P_{\rm jet}$ equal to 10--fold and 100--fold $P_{\rm rad}$.
Same symbols as in Fig. \ref{prld}.
The average error bar is indicated.
}
\end{figure}


\begin{figure}
\includegraphics[height=0.7\textheight]{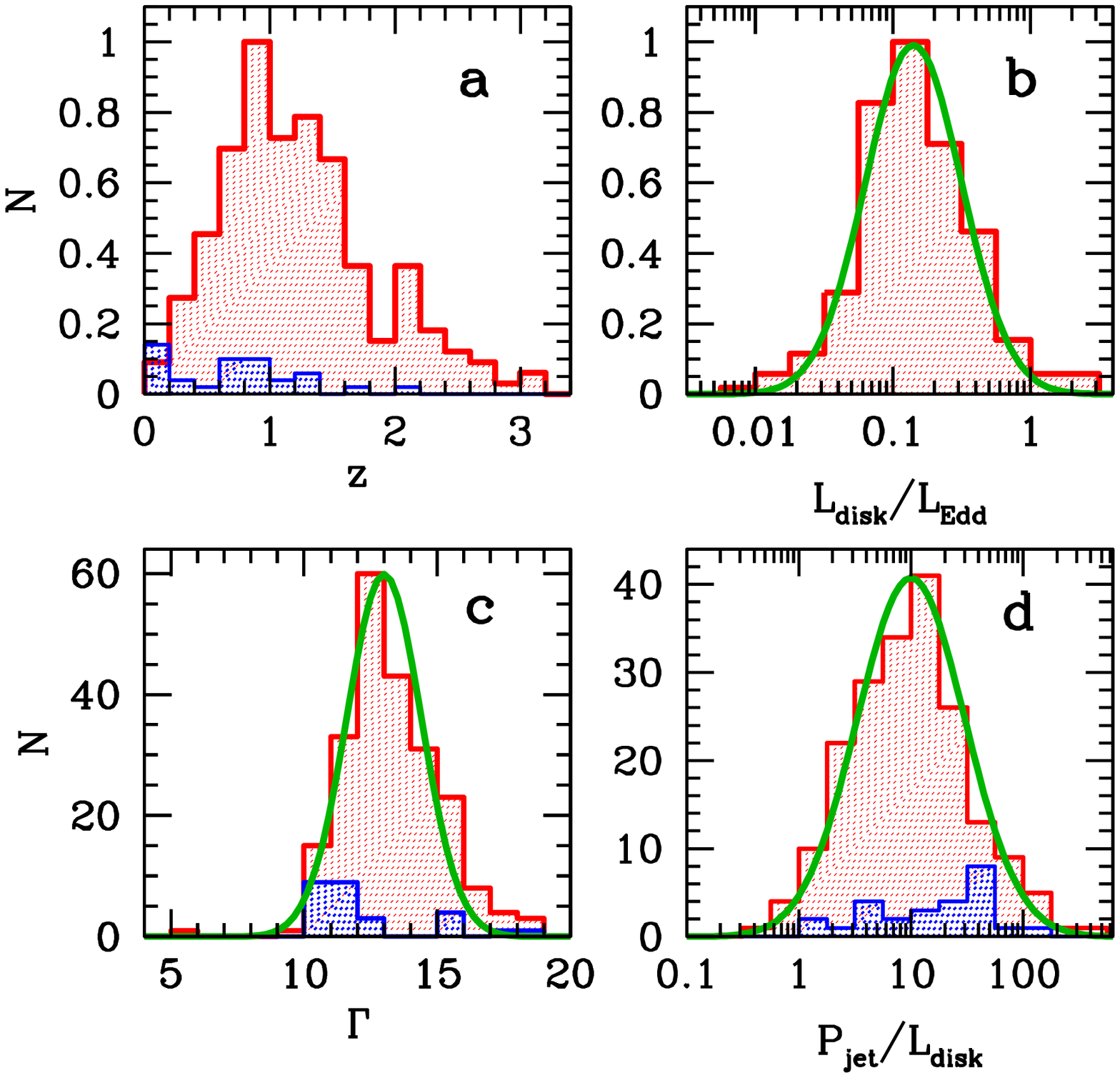}
\caption{
{\bf
Extended Data Figure 2:  Distribution of relevant quantities.
a:} 
Normalized redshift distribution for FSRQ
(red hatching) and BL Lacs (blue hatching) in our sample. 
{\bf b:}
Normalized distribution of the ratio $\log L_{\rm disk}/L_{\rm Eddd}$ for the 
FSRQs in our sample. 
The black hole mass is the virial mass, calculated on the basis of the width of 
the broad lines{\cite{shaw12}}, compared to 
a log--normal distribution having a width $\sigma=0.35$ dex. 
{\bf c:}
Distribution of the bulk Lorentz factor. 
The darker hatched histogram corresponds to BL Lacs.
The plotted gaussian distribution has a width $\sigma=1.4$. 
{\bf d:}
Distribution of the ratio $\log P_{\rm jet}/L_{\rm disk}$ for our sources,
including BL Lacs (darker hatched histogram).
The shown log--normal distribution has a width $\sigma=0.48$ dex.
}
\end{figure}


\begin{figure}
\includegraphics[height=0.7\textheight]{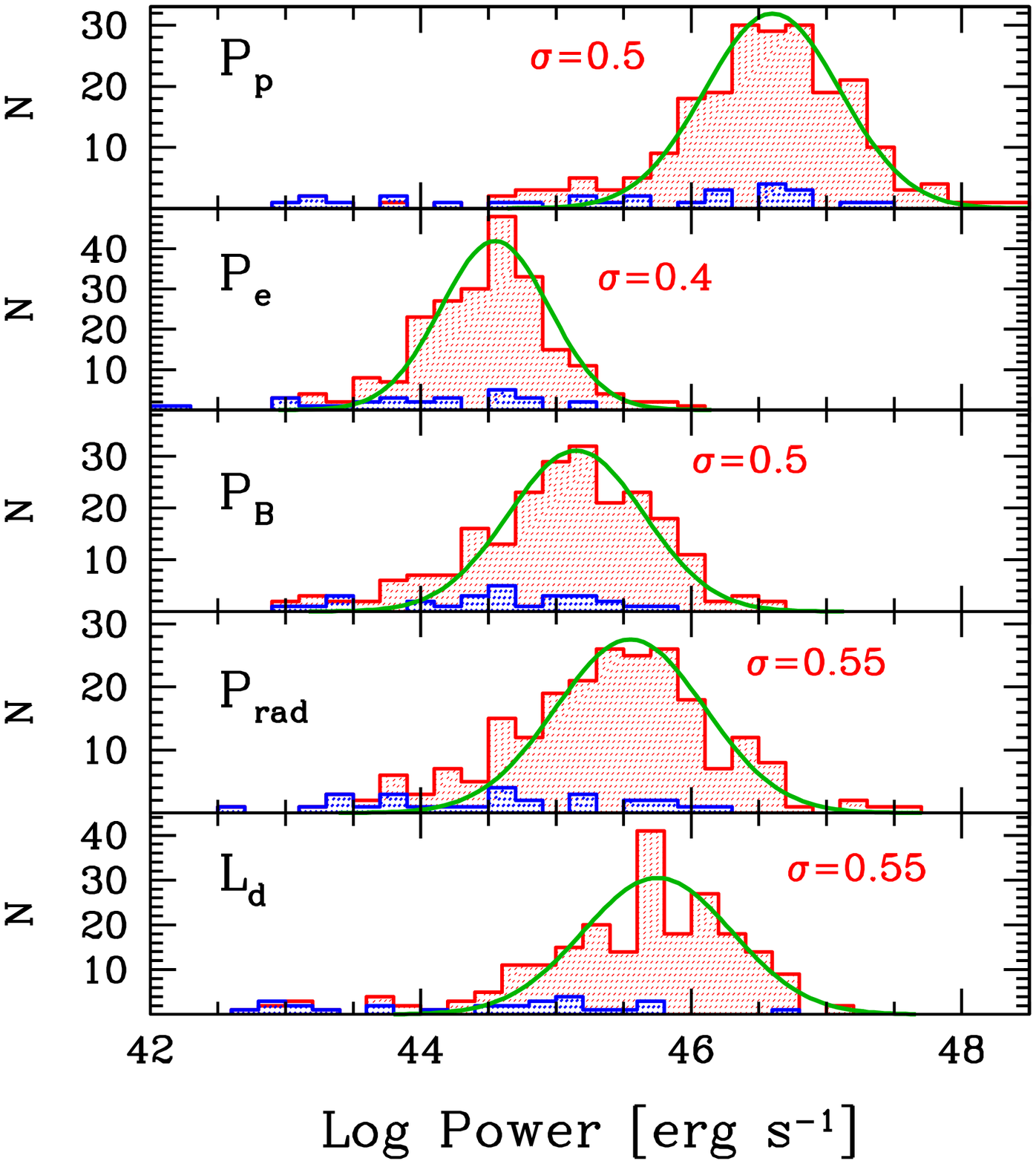}
\caption{
{\bf
Extended Data Figure 3:  Distribution of jet powers.
}
Jet power distribution for FSRQ and BL Lacs 
(darker hatched histogram) in our sample, compared with the
disk luminosity distribution as labelled:
$P_{\rm p}$ is the kinetic power of the (cold) protons, assuming on proton
per emitting electron;
$P_{\rm e}$ is the power in relativistic emitting electrons;
$P_{\rm B}$ is the jet Poynting flux;
$P_{\rm rad}$ is the power that the jet has spent to produce the 
observed radiation;
$L_{\rm disk}$ is the luminosity of the accretion disk.
All distributions are fitted with a log--normal distribution.
The corresponding value of $\sigma$ (in dex) is reported.
The average values of the distributions are:
$\langle \log L_{\rm disk}\rangle = 45.5 $,  
$\langle \log P_{\rm rad}\rangle = 45.3 $, 
$\langle \log P_{\rm B}\rangle = 45.0 $,
$\langle \log P_{\rm e}\rangle = 44.4 $,
$\langle \log P_{\rm p}\rangle = 46.4 $
(units are erg s$^{-1}$).
}
\end{figure}


\begin{table*} 
\centering
\begin{tabular}{llllllllllllllllll}
\hline
\hline
Name &$z$  &$P_{\rm rad}$ &$P_{\rm e}$ &$P_{\rm B}$  &$P_{\rm p}$ &$\Gamma$ &$\theta_{\rm v}$ 
&$L_{\rm disk, fit}$ &$L_{\rm disk, l}$ &$M_{{\rm H}\beta}$ &$M_{\rm MgII}$ &$M_{\rm CIV}$ \\
~[1]     &[2]   &[3]   &[4]   &[5]   &[6]   &[7]   &[8]   &[9]   &[10]   &[11]   &[12]   &[13]    \\
\hline 
\hline
0004 --4736  &0.880  &44.64  &43.98  &44.93  &45.79  &14.0   &2.0  &45.32  &45.11   &---  &7.85  &---  \\   
0011  +0057  &1.493  &45.50  &44.80  &44.87  &47.27  &13.0   &3.0  &45.56  &45.60   &---  &7.80  &7.09  \\
0015  +1700  &1.709  &45.58  &44.16  &45.66  &46.45  &12.0   &3.0  &46.35  &46.26   &---  &9.36  &9.15  \\
0017 --0512  &0.226  &43.98  &44.04  &43.87  &46.49  &12.0   &3.0  &44.65  &45.30   &7.55 &---   &---    \\ 
0023  +4456  &2.023  &45.85  &44.88  &45.16  &46.75  &15.0   &3.0  &45.43  &45.28   &---  &---   &7.78  \\
0024  +0349  &0.545  &43.94  &43.17  &44.38  &44.76  &13.0   &3.0  &44.78  &44.79   &---  &7.76  &---    \\   
0042  +2320  &1.426  &45.74  &44.96  &45.01  &47.20  &13.0   &3.0  &45.62  &45.62   &---  &9.01  &---   \\    
0043  +3426  &0.966  &45.05  &44.05  &44.31  &45.72  &14.0   &3.0  &45.00  &45.02   &---  &7.01  &---  \\     
0044 --8422  &1.032  &45.48  &44.72  &45.27  &47.00  &14.0   &3.0  &45.88  &45.88   &---  &7.68  &---   \\    
0048  +2235  &1.161  &45.46  &44.53  &44.56  &46.70  &13.0   &3.0  &45.26  &45.26   &---  &7.43  &7.25  \\
0050 --0452  &0.922  &44.89  &44.39  &44.74  &46.57  &11.0   &3.0  &45.32  &45.35   &---  &7.20  &---   \\    
0058  +3311  &1.369  &45.15  &44.43  &44.48  &46.48  &13.0   &3.0  &45.18  &45.21   &---  &7.01  &7.97  \\
0102  +4214  &0.874  &45.25  &44.45  &45.09  &46.63  &12.0   &3.0  &45.78  &45.83   &7.92 &7.49  &---  \\     
0102  +5824  &0.644  &45.03  &44.40  &45.26  &46.66  &11.0   &3.0  &45.65  &45.66   &---  &7.57  &---   \\    
0104 --2416  &1.747  &45.63  &44.59  &45.39  &46.51  &11.0   &3.0  &45.91  &45.94   &---  &7.85  &7.97  \\
0157 --4614  &2.287  &45.99  &44.71  &45.08  &46.62  &14.0   &3.0  &45.78  &45.73   &---  &7.98  &7.52  \\
0203  +3041  &0.955  &45.08  &44.66  &43.79  &46.49  &13.0   &3.0  &44.48  &44.40   &---  &7.02  &---      \\ 
0217 --0820  &0.607  &44.34  &43.98  &43.78  &45.98  &13.0   &3.0  &44.20  &44.06   &---  &6.53  &---      \\ 
0226  +0937  &2.605  &45.59  &44.03  &45.93  &46.12  &12.0   &3.0  &46.62  &46.53   &---  &---   &9.65  \\
0237  +2848  &1.206  &45.70  &44.69  &45.68  &47.00  &13.0   &3.0  &46.26  &46.38   &---  &9.22  &---    \\   
0245  +2405  &2.243  &46.29  &44.95  &45.63  &47.40  &14.0   &3.0  &46.32  &46.33   &---  &9.02  &9.18  \\
0246 --4651  &1.385  &46.06  &44.94  &46.08  &47.08  &12.0   &3.0  &46.38  &46.42   &---  &7.48  &7.32  \\
0252 --2219  &1.419  &45.92  &44.90  &45.20  &47.16  &12.0   &3.0  &45.50  &45.72   &---  &9.40  &---       \\
0253  +5102  &1.732  &46.06  &44.66  &45.33  &46.78  &14.0   &3.0  &46.02  &45.99   &---  &9.11  &7.37  \\
0257 --1212  &1.391  &45.27  &44.15  &45.66  &46.32  &11.0   &3.0  &46.35  &46.13   &---  &9.22  &---      \\ 
0303 --6211  &1.348  &45.77  &44.57  &46.26  &46.87  &11.0   &3.0  &46.48  &46.64   &---  &9.76  &---      \\ 
0309 --6058  &1.479  &46.06  &44.83  &45.48  &46.80  &13.0   &3.0  &45.88  &45.88   &---  &7.87  &---      \\ 
0315 --1031  &1.565  &45.44  &44.62  &44.93  &46.89  &13.0   &3.0  &45.62  &45.67   &---  &7.17  &7.33  \\
0325  +2224  &2.066  &46.32  &45.06  &46.09  &47.24  &12.0   &3.0  &46.78  &46.79   &---  &9.50  &9.16  \\
0325 --5629  &0.862  &44.58  &43.92  &44.94  &46.23  &10.0   &3.0  &45.62  &45.60   &---  &7.68  &---       \\
0407  +0742  &1.133  &45.31  &44.56  &45.38  &46.83  &12.0   &3.0  &45.78  &45.51   &---  &7.65  &---       \\
0413 --5332  &1.024  &45.25  &44.66  &44.49  &46.66  &12.0   &3.0  &45.18  &45.14   &---  &7.83  &---       \\
0422 --0643  &0.242  &43.73  &43.83  &44.10  &46.27  &12.0   &3.0  &44.49  &44.42   &7.47 &---   &---     \\
0430 --2507  &0.516  &43.77  &43.15  &43.56  &44.76  &11.0   &3.0  &43.95  &43.81   &---  &6.51  &---       \\
0433  +3237  &2.011  &45.35  &44.22  &45.37  &46.88  &13.0   &3.0  &46.56  &46.58   &---  &9.17  &9.20  \\
0438 --1251  &1.285  &45.09  &44.35  &45.11  &46.27  &11.0   &3.0  &45.80  &45.78   &---  &7.66  &---       \\
0442 --0017  &0.845  &45.71  &44.59  &45.09  &46.45  &13.0   &3.0  &45.78  &45.81   &7.74 &7.46  &---       \\
0448  +1127  &1.370  &45.87  &44.55  &45.39  &46.96  &16.0   &3.0  &46.38  &46.71   &---  &9.44  &---       \\
0449  +1121  &2.153  &45.90  &45.16  &45.24  &46.71  &13.0   &2.5  &45.86  &45.92   &---  &---   &7.89  \\
0456 --3136  &0.865  &44.84  &44.20  &44.79  &46.34  &13.0   &3.0  &45.48  &45.26   &7.78 &7.61  &---       \\
\hline
\hline 
\end{tabular}
\caption{
{\bf Extended Data Table 1: Relevant parameters of the blazars
studied in this paper.
}
Col. 1 and Col. 2 : AR and Dec (J2000);
Col. 2: redshift;
Col. 3 -- Col. 6:
Logarithm of $P_{\rm rad}$,  $P_{\rm e}$, $P_{\rm B}$, $P_{\rm p}$ (powers
in units of erg s$^{-1}$);
Col. 7: bulk Lorentz factor;
Col. 8: viewing angle in degrees;
Col. 9: Logarithm of the disk luminosity (in units of erg s$^{-1}$) as resulting from disk fitting;
Col. 10: Logarithm of the disk luminosity (in units of erg s$^{-1}$) as measured from the broad emission lines;
Col. 11 -- Col. 13: logarithm of the black hole mass (in units of the solar mass)
estimated through the virial method\cite{shaw12} using
the H$\beta$ (Col. 11), MgII (Col. 12) and CIV (Col. 13) broad emission lines.
}
\label{para}
\end{table*}

\setcounter{table}{0}
\begin{table*} 
\centering
\begin{tabular}{llllllllllllllllll}
\hline
\hline
Name &$z$  &$P_{\rm rad}$ &$P_{\rm e}$ &$P_{\rm B}$  &$P_{\rm p}$ &$\Gamma$ &$\theta_{\rm v}$ 
&$L_{\rm disk, fit}$ &$L_{\rm disk, line}$ &$M_{{\rm H}\beta}$ &$M_{\rm MgII}$ &$M_{\rm CIV}$ \\
~[1]     &[2]   &[3]   &[4]   &[5]   &[6]   &[7]   &[8]   &[9]   &[10]   &[11]   &[12]   &[13]    \\
\hline 
\hline
0507 --6104  &1.089  &45.22  &44.82  &45.14  &47.11  &12.0   &3.0  &45.83  &45.86   &---  &7.74  &---       \\
0509  +1011  &0.621  &44.66  &43.88  &44.64  &45.52  &12.0   &3.0  &45.26  &45.35   &7.03 &7.52  &---       \\
0516 --6207  &1.300  &45.52  &44.34  &45.04  &46.03  &12.0   &3.0  &45.73  &45.74   &---  &7.93  &7.52  \\
0526 --4830  &1.300  &45.45  &44.54  &45.09  &46.69  &14.0   &3.0  &45.78  &45.87   &---  &9.15  &7.46  \\
0532  +0732  &1.254  &46.02  &44.95  &45.26  &46.94  &14.0   &3.0  &45.95  &45.86   &---  &7.43  &---       \\
0533  +4822  &1.160  &45.68  &44.32  &46.08  &46.38  &13.0   &3.0  &46.30  &46.26   &---  &9.25  &---       \\
0533 --8324  &0.784  &44.66  &44.06  &44.04  &46.02  &12.0   &3.0  &44.73  &44.73   &---  &7.40  &---       \\
0541 --0541  &0.838  &45.32  &44.51  &45.33  &46.87  &13.0   &3.0  &46.02  &46.06   &---  &7.74  &---       \\
0601 --7036  &2.409  &46.28  &45.27  &45.06  &46.32  &14.0   &3.0  &45.78  &45.69   &---  &---   &7.36  \\
0607 --0834  &0.870  &45.32  &44.45  &45.54  &46.79  &12.0   &3.0  &46.23  &46.33   &7.63 &9.02  &---       \\
0608 --1520  &1.094  &45.42  &44.54  &44.86  &46.75  &14.0   &3.0  &45.56  &45.51   &---  &7.09  &---       \\
0609 --0615  &2.219  &46.60  &44.87  &45.82  &47.26  &16.0   &3.0  &46.52  &46.53   &---  &---   &7.89  \\
0625 --5438  &2.051  &46.58  &44.50  &45.57  &47.30  &15.0   &3.0  &46.56  &46.21   &---  &7.40  &9.07  \\
0645  +6024  &0.832  &45.00  &44.18  &45.63  &46.54  &12.0   &3.0  &46.02  &46.09   &9.09  &9.56  &---       \\
0654  +5042  &1.253  &45.37  &44.22  &45.18  &45.98  &12.0   &3.0  &44.98  &44.97   &---  &7.86  &7.79  \\
0654  +4514  &0.928  &45.43  &44.62  &44.67  &46.35  &12.0   &3.0  &45.26  &45.25   &---  &7.17  &---       \\
0713  +1935  &0.540  &44.39  &43.59  &44.38  &45.09  &12.0   &2.4  &44.78  &44.93   &7.33 &7.91  &---       \\
0721  +0406  &0.665  &45.23  &43.96  &45.87  &46.07  &13.0   &4.0  &46.26  &46.33   &7.49 &9.12  &---       \\
0723  +2859  &0.966  &44.74  &43.66  &45.38  &45.53  &12.0   &3.0  &45.77  &45.75   &---  &7.40  &---       \\
0725  +1425  &1.038  &45.80  &44.60  &45.26  &46.93  &14.0   &3.0  &45.95  &45.95   &---  &7.31  &---       \\
0746  +2549  &2.979  &47.35  &46.00  &45.78  &48.40  &16.0   &3.0  &46.48  &46.31   &---  &---   &9.23  \\
0805  +6144  &3.033  &47.18  &45.85  &45.83  &48.16  &15.0   &3.0  &46.52  &46.56   &---  &---   &9.07  \\
0825  +5555  &1.418  &46.02  &44.90  &45.66  &47.33  &14.0   &3.0  &46.35  &46.32   &---  &9.10  &---       \\
0830  +2410  &0.942  &45.82  &44.80  &45.96  &47.01  &12.0   &3.0  &46.35  &45.97   &---  &7.70  &---       \\
0840  +1312  &0.680  &44.83  &44.39  &45.04  &46.86  &12.0   &3.0  &45.73  &45.75   &7.37 &7.62  &---       \\
0847 --2337  &0.059  &43.67  &43.22  &43.24  &44.98  &5.0    &10.0 &43.18  &43.32   &7.30 &---   &---     \\
0856  +2111  &2.098  &45.73  &44.12  &45.63  &46.24  &15.0   &3.0  &46.42  &47.11   &---  &9.96  &9.77  \\
0909  +0121  &1.026  &45.96  &44.76  &46.01  &47.25  &13.0   &3.0  &46.53  &46.47   &---  &9.14  &---       \\
0910  +2248  &2.661  &46.37  &45.04  &45.48  &47.19  &16.0   &3.0  &46.18  &46.21   &---  &---   &7.70  \\
0912  +4126  &2.563  &45.99  &44.66  &46.05  &46.89  &11.0   &3.0  &46.35  &46.36   &---  &---   &9.32  \\
0920  +4441  &2.189  &46.64  &45.30  &46.41  &47.81  &12.0   &3.0  &46.71  &46.70   &---  &---   &9.29  \\
0921  +6215  &1.453  &45.72  &44.65  &45.71  &46.71  &12.0   &3.0  &46.01  &46.05   &---  &7.93  &---       \\
0923  +2815  &0.744  &44.92  &44.14  &44.79  &45.91  &13.0   &3.0  &45.48  &45.63   &7.61 &9.04  &---       \\
0923  +4125  &1.732  &45.47  &44.51  &44.18  &45.98  &13.0   &3.0  &44.88  &44.75   &---  &7.68  &7.16  \\
0926  +1451  &0.632  &44.61  &43.97  &43.09  &45.30  &14.0   &3.0  &43.78  &43.75   &---  &7.11  &---       \\
0937  +5008  &0.276  &43.64  &43.96  &43.58  &46.18  &11.0   &3.0  &43.78  &43.99   &7.50 &---   &---     \\
0941  +2778  &1.305  &45.02  &43.98  &45.02  &45.83  &12.0   &3.0  &45.71  &45.72   &---  &7.63  &---       \\
0946  +1017  &1.006  &45.20  &44.05  &44.98  &46.05  &12.0   &3.0  &45.67  &45.74   &---  &7.47  &---       \\
0948  +0022  &0.585  &45.01  &44.66  &44.79  &47.22  &15.0   &3.0  &45.35  &45.03   &7.46 &7.73  &---       \\
0949  +1752  &0.693  &44.92  &44.33  &44.55  &46.53  &14.0   &3.0  &45.25  &45.25   &---  &7.10  &---       \\
0956  +2515  &0.708  &44.96  &44.32  &45.23  &46.60  &10.0   &3.0  &45.88  &45.93   &7.30 &7.63  &---       \\
0957  +5522  &0.899  &45.87  &44.76  &44.87  &46.84  &11.0   &4.0  &45.65  &45.59   &---  &7.45  &---       \\
\hline
\hline 
\end{tabular}
\caption{ {\bf Extended Data Table 1: Relevant parameters of the blazars
studied in this paper.
} 
{\it Continue.}
}
\label{para}
\end{table*}

\setcounter{table}{0}
\begin{table*} 
\centering
\begin{tabular}{llllllllllllllllll}
\hline
\hline
Name &$z$  &$P_{\rm rad}$ &$P_{\rm e}$ &$P_{\rm B}$  &$P_{\rm p}$ &$\Gamma$ &$\theta_{\rm v}$ 
&$L_{\rm disk, fit}$ &$L_{\rm disk, line}$ &$M_{{\rm H}\beta}$ &$M_{\rm MgII}$ &$M_{\rm CIV}$ \\
~[1]     &[2]   &[3]   &[4]   &[5]   &[6]   &[7]   &[8]   &[9]   &[10]   &[11]   &[12]   &[13]    \\
\hline 
\hline
1001  +2911  &0.558  &44.57  &44.73  &43.86  &46.46  &16.0   &3.0  &43.95  &44.06   &7.31 &7.64  &---       \\
1012  +2439  &1.800  &45.77  &44.28  &44.81  &46.24  &11.0   &3.0  &46.13  &45.56   &---  &7.73  &7.86  \\
1016  +0513  &1.714  &45.88  &44.61  &45.26  &46.53  &12.0   &3.0  &45.65  &45.55   &---  &7.34  &7.64  \\
1018  +3542  &1.228  &46.46  &44.92  &45.66  &47.29  &15.0   &3.0  &46.35  &46.34   &---  &9.10  &---       \\
1022  +3931  &0.604  &44.49  &43.57  &45.19  &45.46  &11.0   &3.0  &45.88  &45.89   &---  &7.95  &---       \\
1032  +6051  &1.064  &45.03  &44.14  &44.69  &46.39  &14.0   &3.0  &45.38  &45.35   &---  &7.75  &---       \\
1033  +4116  &1.117  &45.45  &44.39  &45.26  &46.13  &14.0   &3.0  &45.78  &45.92   &---  &7.61  &---       \\
1033  +6051  &1.401  &46.04  &45.11  &45.09  &47.31  &15.0   &3.0  &45.71  &45.66   &---  &9.09  &---       \\
1037  +2823  &1.066  &45.47  &44.34  &45.34  &46.28  &12.0   &3.0  &46.03  &45.95   &---  &7.99  &---       \\
1043  +2408  &0.559  &44.17  &43.88  &43.72  &46.06  &9.0    &3.0  &44.68  &44.65   &---  &7.09  &---       \\
1058  +0133  &0.888  &45.61  &44.68  &45.20  &46.54  &14.0   &3.0  &45.48  &45.51   &---  &7.37  &---       \\
1106  +2812  &0.843  &44.90  &44.87  &44.76  &46.50  &11.0   &3.0  &45.20  &46.26   &---  &7.85  &---       \\
1112  +3446  &1.956  &45.89  &44.60  &45.59  &46.52  &10.0   &3.0  &46.28  &46.13   &---  &7.74  &7.82  \\
1120  +0704  &1.336  &45.01  &44.59  &44.26  &46.16  &12.0   &3.0  &44.95  &45.47   &---  &7.83  &---       \\
1124  +2336  &1.549  &45.45  &44.48  &45.25  &46.35  &13.0   &3.0  &45.83  &46.05   &---  &7.79  &---       \\
1133  +0040  &1.633  &45.75  &44.42  &45.18  &46.46  &14.0   &3.0  &45.88  &45.64   &---  &7.80  &---       \\
1146  +3958  &1.088  &45.57  &44.53  &45.63  &46.54  &12.0   &3.0  &46.07  &46.06   &---  &7.93  &---       \\
1152 --0841  &2.367  &45.98  &44.88  &46.05  &47.28  &13.0   &3.0  &46.28  &46.25   &---  &---   &9.38  \\
1154  +6022  &1.120  &45.59  &44.52  &45.26  &46.53  &13.0   &3.0  &45.95  &45.97   &---  &7.94  &---       \\
1155 --8101  &1.395  &45.74  &44.71  &44.78  &46.65  &14.0   &3.0  &45.48  &45.55   &---  &7.30  &---       \\
1159  +2914  &0.725  &45.45  &44.52  &44.87  &46.33  &12.0   &3.0  &45.56  &45.65   &7.14 &7.61  &---       \\
1208  +5441  &1.344  &45.88  &44.81  &44.93  &46.74  &14.0   &3.0  &45.62  &45.53   &---  &7.40  &---       \\
1209  +1810  &0.845  &44.80  &43.91  &44.99  &45.83  &12.0   &3.0  &45.68  &45.47   &7.26 &7.77  &---       \\
1222  +0413  &0.966  &45.94  &45.11  &45.78  &47.34  &12.0   &3.0  &46.18  &45.97   &---  &7.37  &---       \\
1224  +2122  &0.434  &45.49  &44.18  &45.39  &46.26  &13.0   &3.0  &46.08  &46.16   &7.89 &7.91  &---       \\
1224  +5001  &1.065  &45.74  &44.44  &45.86  &46.66  &14.0   &3.0  &46.56  &45.85   &---  &7.66  &---       \\
1226  +4340  &2.001  &46.41  &44.65  &45.42  &47.09  &16.0   &3.0  &46.41  &46.14   &---  &7.64  &9.01  \\
1228  +4858  &1.722  &45.61  &44.43  &45.07  &46.19  &12.0   &3.0  &45.65  &45.68   &---  &7.28  &7.23  \\
1239  +0443  &1.761  &46.81  &45.48  &45.14  &47.58  &15.0   &3.0  &45.83  &45.83   &---  &7.46  &7.68  \\
1257  +3229  &0.806  &45.04  &44.38  &45.40  &46.71  &13.0   &3.0  &45.62  &45.28   &7.89 &7.62  &---       \\
1303 --4621  &1.664  &45.32  &44.57  &44.39  &46.69  &13.0   &3.0  &45.18  &45.21   &---  &7.95  &7.21  \\
1310  +3220  &0.997  &45.53  &44.62  &45.27  &46.86  &11.0   &3.0  &45.95  &45.92   &---  &7.57  &---       \\
1317  +3425  &1.055  &44.86  &44.14  &45.81  &46.52  &11.0   &3.0  &46.03  &46.09   &---  &9.14  &---       \\
1321  +2216  &0.943  &45.03  &44.30  &45.16  &45.58  &13.0   &3.0  &45.38  &45.99   &7.87 &7.76  &---       \\
1327  +2210  &1.403  &45.97  &45.02  &45.41  &47.29  &13.0   &3.0  &46.02  &45.96   &---  &9.25  &---       \\
1332 --1256  &1.492  &46.14  &44.57  &45.56  &46.92  &15.0   &3.0  &46.26  &46.26   &---  &7.96  &7.61  \\
1333  +5057  &1.362  &45.53  &44.37  &44.69  &46.09  &15.0   &3.0  &45.38  &45.36   &---  &7.95  &---       \\
1343  +5754  &0.933  &45.13  &44.14  &45.09  &46.40  &12.0   &4.0  &45.78  &45.65   &---  &7.42  &---       \\
1344 --1723  &2.506  &46.19  &44.84  &45.64  &45.94  &17.0   &2.4  &46.03  &46.02   &---  &---   &9.12  \\
1345  +4452  &2.534  &46.60  &45.24  &45.62  &47.28  &14.0   &3.0  &46.02  &46.12   &---  &---   &7.98  \\
1347 --3750  &1.300  &45.34  &44.69  &45.04  &47.00  &12.0   &3.0  &45.73  &45.67   &---  &7.95  &7.62  \\
1350  +3034  &0.712  &44.83  &44.33  &44.79  &46.53  &12.0   &3.0  &45.54  &45.28   &7.21 &7.33  &---       \\
1357  +7643  &1.585  &45.57  &44.84  &44.48  &46.84  &15.0   &3.0  &45.18  &45.20   &---  &7.34  &7.17  \\
1359  +5544  &1.014  &45.30  &44.53  &44.44  &46.20  &13.0   &3.0  &45.13  &44.99   &---  &7.00  &---       \\
\hline
\hline 
\end{tabular}
\caption{ {\bf Extended Data Table 1: Relevant parameters of the blazars
studied in this paper.
} 
{\it Continue.}
}
\label{para}
\end{table*}

\setcounter{table}{0}
\begin{table*} 
\centering
\begin{tabular}{llllllllllllllllll}
\hline
\hline
Name &$z$  &$P_{\rm rad}$ &$P_{\rm e}$ &$P_{\rm B}$  &$P_{\rm p}$ &$\Gamma$ &$\theta_{\rm v}$ 
&$L_{\rm disk, fit}$ &$L_{\rm disk, line}$ &$M_{{\rm H}\beta}$ &$M_{\rm MgII}$ &$M_{\rm CIV}$ \\
~[1]     &[2]   &[3]   &[4]   &[5]   &[6]   &[7]   &[8]   &[9]   &[10]   &[11]   &[12]   &[13]    \\
\hline 
\hline
1423 --7829  &0.788  &44.59  &43.96  &44.56  &46.45  &12.0   &3.0  &45.26  &45.29   &7.14 &7.32  &---       \\
1436  +2321  &1.548  &45.39  &44.40  &45.77  &46.36  &11.0   &3.0  &46.17  &45.66   &---  &7.12  &7.50  \\
1438  +3710  &2.399  &46.68  &45.74  &45.48  &47.34  &17.0   &3.0  &46.18  &46.36   &---  &---   &7.58  \\
1439  +3712  &1.027  &45.45  &44.07  &45.88  &46.36  &14.0   &3.0  &46.57  &46.16   &---  &9.08  &---       \\
1441 -1523  &2.642  &46.55  &45.54  &45.50  &47.90  &12.0   &3.0  &46.19  &46.20   &---  &---   &7.49  \\
1443  +2501  &0.939  &44.64  &44.12  &45.03  &46.72  &10.0   &3.0  &45.26  &45.28   &7.42 &7.84  &---       \\
1504  +1029  &1.839  &47.27  &45.31  &45.66  &47.28  &18.0   &2.0  &46.18  &46.17   &---  &7.98  &7.90  \\
1505  +0326  &0.409  &44.49  &43.95  &44.14  &45.94  &14.0   &3.0  &44.83  &44.72   &---  &7.41  &---       \\
1514  +4450  &0.570  &44.30  &44.10  &44.07  &46.32  &11.0   &3.0  &44.30  &44.33   &7.72 &7.62  &---       \\
1522  +3144  &1.484  &45.90  &44.72  &44.96  &47.03  &12.0   &3.0  &45.65  &45.90   &---  &7.92  &---       \\
1539  +2744  &2.191  &45.66  &44.32  &45.26  &46.04  &12.0   &3.0  &45.65  &45.63   &---  &7.43  &7.51  \\
1549  +0237  &0.414  &44.59  &43.92  &45.09  &46.12  &11.0   &3.0  &45.78  &45.83   &7.62 &7.72  &---       \\
1550  +0527  &1.417  &45.59  &44.78  &45.53  &47.07  &12.0   &3.0  &46.08  &46.08   &---  &7.98  &---       \\
1553  +1256  &1.308  &45.58  &43.91  &46.09  &45.97  &12.0   &3.0  &46.78  &46.18   &---  &7.64  &---       \\
1608  +1029  &1.232  &45.63  &44.77  &45.86  &47.17  &12.0   &3.0  &46.26  &46.07   &---  &7.77  &---       \\
1613  +3412  &1.400  &45.29  &44.41  &46.43  &46.96  &10.0   &3.0  &46.73  &46.46   &---  &9.08  &---       \\
1616  +4632  &0.950  &45.21  &44.29  &44.78  &46.49  &14.0   &3.0  &45.48  &45.42   &---  &7.28  &---       \\
1617 --5848  &1.422  &46.40  &44.88  &46.39  &47.36  &15.0   &3.0  &47.02  &47.00   &---  &9.81  &9.01  \\
1624 --0649  &3.037  &45.96  &44.53  &45.86  &47.01  &14.0   &3.0  &46.56  &46.35   &---  &---   &7.23  \\
1628 --6152  &2.578  &46.11  &45.05  &45.63  &47.33  &12.0   &3.0  &46.02  &46.03   &---  &---   &7.92  \\
1635  +3808  &1.813  &47.69  &45.58  &46.56  &48.04  &14.0   &3.0  &46.78  &46.66   &---  &9.30  &7.85  \\
1636  +4715  &0.823  &45.73  &44.70  &44.86  &47.01  &15.0   &3.0  &45.56  &45.49   &7.11 &7.38  &---       \\
1637  +4717  &0.735  &45.26  &44.38  &44.86  &46.09  &12.0   &3.0  &45.26  &45.58   &7.61 &7.52  &---       \\
1639  +4705  &0.860  &45.53  &44.21  &45.26  &46.33  &14.0   &3.0  &45.95  &45.97   &---  &7.95  &---       \\
1642  +3940  &0.593  &45.67  &44.83  &45.73  &46.90  &11.0   &3.5  &45.95  &46.01   &7.73 &9.03  &---       \\
1703 --6212  &1.747  &46.37  &45.06  &46.03  &47.22  &12.0   &3.0  &46.72  &46.31   &---  &7.65  &7.55  \\
1709  +4318  &1.027  &45.53  &44.64  &44.63  &46.32  &15.0   &3.0  &45.18  &45.03   &---  &7.92  &---       \\
1734  +3857  &0.975  &45.40  &44.27  &44.91  &46.00  &13.0   &3.0  &45.30  &45.01   &---  &7.97  &---       \\
1736  +0631  &2.387  &46.45  &45.16  &45.39  &47.32  &15.0   &3.0  &46.08  &46.21   &---  &7.82  &9.39  \\
1802 --3940  &1.319  &46.24  &44.78  &45.66  &46.78  &12.0   &3.0  &46.18  &46.11   &---  &7.60  &7.59  \\
1803  +0341  &1.420  &45.90  &44.92  &44.39  &46.53  &15.0   &3.0  &45.08  &45.01   &---  &7.79  &---       \\
1818  +0903  &0.354  &44.11  &43.43  &44.19  &45.01  &12.0   &3.0  &44.88  &44.93   &7.30 &7.50  &---       \\
1830  +0619  &0.745  &45.22  &44.16  &45.96  &46.66  &13.0   &3.0  &46.35  &46.45   &7.69 &7.86  &---       \\
1848  +3219  &0.800  &45.27  &44.61  &45.26  &46.89  &13.0   &3.0  &45.65  &45.58   &7.87 &7.21  &---       \\
1903 --6749  &0.254  &43.84  &44.00  &44.10  &46.36  &11.0   &3.0  &44.35  &44.42   &7.51 &---   &---     \\
1916 --7946  &0.204  &44.14  &43.53  &43.95  &45.20  &13.0   &3.0  &44.73  &44.90   &7.82 &---   &---     \\
1928 --0456  &0.587  &45.45  &44.36  &44.71  &46.40  &15.0   &3.0  &45.62  &45.62   &---  &9.07  &---       \\
1954 --1123  &0.683  &44.80  &44.04  &43.90  &45.46  &11.0   &3.0  &44.48  &44.37   &---  &6.73  &---       \\
1955  +1358  &0.743  &45.20  &44.59  &45.26  &46.82  &11.0   &3.0  &45.65  &45.73   &7.17 &7.39  &---       \\
1959 --4246  &2.178  &46.03  &44.61  &45.86  &46.62  &12.0   &3.0  &46.16  &46.13   &---  &7.55  &9.41  \\
2017  +0603  &1.743  &46.06  &44.46  &46.12  &46.84  &15.0   &3.0  &47.20  &47.42   &---  &9.39  &9.67  \\
2025 --2845  &0.884  &45.73  &44.86  &44.36  &46.67  &13.0   &3.0  &45.05  &45.01   &---  &7.34  &---       \\
2031  +1219  &1.213  &45.20  &44.86  &44.48  &45.93  &18.0   &2.3  &44.88  &44.76   &---  &7.99  &7.19  \\
2035  +1056  &0.601  &44.94  &44.30  &44.78  &46.37  &12.0   &3.0  &45.18  &45.11   &7.74 &7.26  &---       \\
\hline
\hline 
\end{tabular}
\caption{ {\bf Extended Data Table 1: Relevant parameters of the blazars
studied in this paper.
} 
{\it Continue.}
}
\label{para}
\end{table*}

\setcounter{table}{0}
\begin{table*} 
\centering
\begin{tabular}{llllllllllllllllll}
\hline
\hline
Name &$z$  &$P_{\rm rad}$ &$P_{\rm e}$ &$P_{\rm B}$  &$P_{\rm p}$ &$\Gamma$ &$\theta_{\rm v}$ 
&$L_{\rm disk, fit}$ &$L_{\rm disk, line}$ &$M_{{\rm H}\beta}$ &$M_{\rm MgII}$ &$M_{\rm CIV}$ \\
~[1]     &[2]   &[3]   &[4]   &[5]   &[6]   &[7]   &[8]   &[9]   &[10]   &[11]   &[12]   &[13]    \\
\hline 
\hline
2110  +0809  &1.580  &45.21  &44.19  &45.65  &46.43  &13.0   &3.0  &46.05  &46.08   &---  &7.82  &---       \\
2118  +0013  &0.463  &44.24  &43.23  &44.17  &45.26  &12.0   &4.0  &44.78  &44.93   &7.60 &7.89  &---       \\
2121  +1901  &2.180  &45.98  &44.74  &45.09  &46.97  &13.0   &3.0  &45.78  &45.26   &---  &---   &7.75  \\
2135 --5006  &2.181  &46.42  &45.12  &45.43  &47.29  &16.0   &3.0  &46.12  &46.36   &---  &7.31  &7.40  \\
2139 --6732  &2.009  &46.56  &45.46  &45.18  &47.79  &15.0   &3.0  &45.88  &45.77   &---  &7.49  &7.93  \\
2145 --3357  &1.361  &44.99  &44.40  &44.77  &46.15  &13.0   &2.0  &45.35  &45.17   &---  &7.31  &---       \\
2157  +3127  &1.448  &45.73  &44.69  &45.38  &46.62  &13.0   &2.4  &45.73  &45.74   &---  &7.89  &---       \\
2202 --8338  &1.865  &45.86  &45.10  &45.88  &47.55  &14.0   &3.0  &46.18  &46.18   &---  &9.02  &9.16  \\
2212  +2355  &1.125  &44.93  &44.26  &45.09  &46.94  &13.0   &3.0  &45.78  &45.78   &---  &7.46  &---       \\
2219  +1806  &1.071  &44.70  &43.76  &44.48  &45.83  &14.0   &3.0  &45.18  &45.07   &---  &7.65  &7.66  \\
2229 --0832  &1.560  &46.32  &45.05  &45.83  &47.17  &13.0   &3.0  &46.48  &46.45   &---  &7.70  &7.54  \\
2236  +2828  &0.790  &45.22  &44.17  &45.58  &46.12  &11.0   &3.0  &45.38  &45.37   &---  &7.35  &---       \\
2237 --3921  &0.297  &45.06  &44.62  &44.26  &46.87  &13.0   &6.0  &44.78  &44.87   &7.77 &7.95  &---       \\
2240  +4057  &1.171  &45.41  &44.67  &44.98  &46.81  &14.0   &2.5  &45.38  &45.31   &---  &7.28  &---       \\
2315 --5018  &0.808  &44.66  &44.38  &44.80  &45.76  &13.0   &3.0  &44.65  &44.62   &---  &7.68  &---       \\
2321  +3204  &1.489  &45.54  &44.50  &45.21  &46.36  &12.0   &3.0  &45.73  &45.71   &---  &7.66  &7.75  \\
2327  +0940  &1.841  &46.42  &45.32  &45.79  &47.76  &12.0   &3.0  &46.48  &46.20   &---  &7.70  &9.35  \\
2331 --2148  &0.563  &44.83  &44.20  &44.04  &45.89  &12.0   &3.0  &44.88  &44.82   &7.53 &7.63  &---       \\
2334  +0736  &0.401  &44.28  &43.60  &45.12  &45.85  &12.0   &3.0  &45.73  &45.76   &7.37 &---   &---     \\
2345 --1555  &0.621  &45.03  &44.15  &44.80  &45.88  &12.0   &3.0  &45.32  &45.30   &7.16 &7.48  &---       \\
2357  +0448  &1.248  &45.69  &44.97  &44.74  &47.16  &15.0   &3.0  &45.43  &46.02   &---  &7.41  &7.45  \\
\hline
\hline
0013  +1907  &0.477  &43.76  &43.51  &43.92  &45.12  &11.0   &3.0  &43.78  &43.70   &---  &---   &---       \\
0203  +3042  &0.761  &44.71  &43.85  &45.87  &46.19  &11.0   &3.0  &45.73  &45.75   &---  &---   &---       \\
0334 --4008  &1.357  &45.74  &45.25  &44.93  &46.83  &17.0   &2.5  &44.86  &44.83   &---  &---   &---       \\
0434 --2014  &0.928  &44.57  &43.85  &44.32  &45.68  &11.0   &3.0  &44.18  &44.15   &---  &---   &---       \\
0438 --4521  &2.017  &45.66  &44.72  &45.03  &46.77  &11.0   &3.0  &45.26  &45.24   &---  &---   &---       \\
0516 --6207  &1.300  &45.65  &44.57  &45.25  &46.70  &11.0   &3.0  &45.48  &45.43   &---  &---   &---       \\
0629 --2001  &1.724  &45.85  &44.71  &45.33  &46.55  &11.0   &3.0  &45.13  &45.05   &---  &---   &---       \\
0831  +0429  &0.174  &43.77  &43.68  &43.14  &44.68  &10.0   &3.0  &43.65  &43.62   &---  &---   &---       \\
1117  +2013  &0.138  &43.30  &42.28  &44.58  &43.10  &18.0   &2.5  &42.95  &42.83   &---  &---   &---       \\
1125 --3559  &0.284  &43.50  &43.02  &44.29  &44.27  &15.0   &3.0  &44.43  &44.34   &---  &---   &---       \\
1203  +6030  &0.065  &42.65  &42.02  &44.45  &43.46  &15.0   &3.0  &43.00  &43.00   &---  &---   &---       \\
1221  +2814  &0.103  &43.36  &43.32  &43.05  &43.78  &10.0   &3.0  &43.08  &43.11   &---  &---   &---       \\
1221  +3010  &0.184  &43.77  &43.02  &43.28  &43.23  &10.0   &3.0  &43.08  &43.05   &---  &---   &---       \\
1420  +5422  &0.153  &44.00  &43.89  &43.33  &44.60  &10.0   &4.0  &43.26  &43.20   &---  &---   &---       \\
1534  +3720  &0.144  &43.38  &42.99  &43.42  &43.12  &11.0   &5.0  &42.78  &42.72   &---  &---   &---       \\
1540  +1438  &0.606  &44.30  &44.23  &44.46  &45.97  &10.0   &3.0  &44.56  &44.57   &---  &---   &---       \\
1755 --6423  &1.255  &45.22  &44.64  &45.25  &46.83  &10.0   &3.0  &45.18  &45.16   &---  &---   &---       \\
1824  +5651  &0.664  &45.18  &44.58  &44.91  &46.67  &10.0   &3.0  &44.91  &44.91   &---  &---   &---       \\
2015  +3709  &0.859  &46.13  &45.17  &45.13  &47.37  &12.0   &3.0  &45.65  &45.18   &---  &---   &---       \\
2152  +1735  &0.874  &44.67  &44.10  &44.79  &44.90  &10.0   &3.0  &45.18  &45.15   &---  &---   &---       \\
2206 --0029  &1.053  &44.72  &44.51  &44.61  &46.24  &11.0   &3.0  &44.83  &44.80   &---  &---   &---       \\
2206  +6500  &1.121  &46.01  &44.74  &45.64  &47.24  &15.0   &3.0  &46.62  &46.67   &---  &---   &---       \\
2236  +2828  &0.790  &45.15  &44.58  &45.32  &46.16  &10.5   &3.0  &45.62  &45.64   &---  &---   &---       \\
2247 --0002  &0.949  &44.69  &44.30  &44.61  &45.45  &11.0   &3.0  &45.13  &45.10   &---  &---   &---       \\
2315 --5018  &0.811  &44.65  &44.00  &44.68  &45.68  &12.0   &3.0  &44.68  &44.62   &---  &---   &---       \\
2353 --3034  &0.737  &44.35  &43.56  &44.56  &45.20  &10.0   &3.0  &44.68  &44.63   &---  &---   &---       \\
\hline
\hline 
\end{tabular}
\caption{ {\bf Extended Data Table 1: Relevant parameters of the blazars
studied in this paper. 
}
{\it Continue.}
The bottom part of this table refer to BL Lac objects. 
}
\label{para}
\end{table*}



\begin{thebibliography}{30} 

\bibitem{blandford78} Blandford, R.D. \& Znajek, R.L.
Electromagnetic extraction of energy from Kerr black holes.
 {\it Mon. Not. R. Astr. Soc.} {\bf 179}, 433--456 (1977)

\bibitem{rs91} Rawlings, S. \& Saunders, R.,   
Evidence for a common central-engine mechanism in all extragalactic radio sources.
{\it Nature} {\bf 439}, 138--140 (1991)

\bibitem{celotti93} Celotti, A. \& Fabian, A.C.   
The Kinetic Power and Luminosity of Parsecscale Radio Jets - an Argument for Heavy Jets.
{\it Mon. Not. R. Astr. Soc.} {\bf 264}, 228--236 (1993)

\bibitem{celotti97} Celotti, A., Padovani, P. \& Ghisellini, G. 
Jets and accretion processes in Active Galactic Nuclei: further clues.
{\it Mon. Not. R. Astr. Soc.} {\bf 286}, 415--424 (1997)

\bibitem{maraschi03} Maraschi, L. \& Tavecchio, F.
The Jet--Disk Connection and Blazar Unification.
{\it Astrophys. J.} {\bf 593}, 667--675 (2003)

\bibitem{punsly06} Punsly, B. \& Tingay, S.J. 
PKS 1018--42: A Powerful, Kinetically Dominated Quasar.
{\it Astrophys. J.} {\bf 640}, L21--L24

\bibitem{celotti08} Celotti, A. \& Ghisellini, G.  
The power of blazar jets.  
{\it Mon. Not. R. Astr. Soc.} {\bf 385}, 283--300 (2008)

\bibitem{gg10general} Ghisellini, G., Tavecchio, F., Foschini, L., Ghirlanda, G., Maraschi, L. \& Celotti, A.
General physical properties of bright Fermi blazars.
{\it Mon. Not. R. Astr. Soc.} {\bf 402}, 497--518  (2010) 

\bibitem{tchekhovskoy11} Tchekhovskoy, A., Narayan, R. \& McKinney, J.C.
Efficient generation of jets from magnetically arrested accretion on a rapidly spinning black hole.
{\it Mon. Not. R. Astr. Soc.} {\bf 418}, L79--L83 (2011)

\bibitem{zamaninasab14} Zamaninasab, M., Clausen--Brown, E., Savolainen, T. \& Tchekhoskoy, A.
        Dynamically important magnetic fields near accreting supermassive black holes.
        {\it Nature} {\bf 512}, 126--128 (2014)  

\bibitem{thorne74} Thorne, K.  
Disk--Accretion onto a Black Hole. II. Evolution of the Hole.
{\it Astrophys. J.} {\bf 191}, 507--519 (1974)

\bibitem{shaw12} Shaw, M.S., Romani, R.W., Cotter, G., et al. 
Spectroscopy of broad--line Blazars from 1LAC.
{\it Astrophys. J.} {\bf 748}, 49 (12pp) (2012)

\bibitem{shaw13} Shaw, M.S., Romani, R.W., Cotter, G. et al. 
Spectroscopy of the Largest Ever $\gamma$--ray-selected BL Lac Sample.
{\it Astrophys. J.} {\bf 764}, 135 (13pp) (2013)

\bibitem{francis91} Francis, J., Hewett, P.C., Foltz, C.B., Chaffee, F.H., Weymann, R.J. \& Morris, S.L.
A high signal--to--noise ratio composite quasar spectrum.
{\it Astrophys. J.}, {\bf 373}, 465--470 (1991)

\bibitem{vanderberk01} 	
Vanden Berk, D.E., Richards, G.T. \& Bauer, A.
Composite Quasar Spectra from the Sloan Digital Sky Survey.
{\it Astron. J.} {\bf 122}, 549--564 (2001)

\bibitem{calderone13} Calderone, G., Ghisellini, G.,  Colpi, M. \& Dotti, M.
Black hole mass estimate for a sample of radio-loud narrow-line Seyfert 1 galaxies.
{\it Mon. Not. R. Astr. Soc.} {\bf 431}, 210--239 (2013) 

\bibitem{gg09} Ghisellini, G. \& Tavecchio, F.  
Canonical high--power blazars.
{\it Mon. Not. R. Astr. Soc.} {\bf 397}, 985--1002 (2009)

\bibitem{gg10} Ghisellini, G. \& Tavecchio, F.
Compton rockets and the minimum power of relativistic jets. 
{\it Mon. Not. R. Astr. Soc.} {\bf 409}, L79--L83 (2010) 

\bibitem{nolan12} Nolan P.L., Abdo A. A., Ackermann M., et al. 
Fermi Large Area Telescope Second Source Catalog.
{\it Astrophys. J. Suppl.} {\bf 199}, 31 (46pp) (2012)

\bibitem{ghirlanda11} Ghirlanda, G., Ghisellini, G., Tavecchio, F., Foschini, L. \& Bonnoli, G.
The radio--$\gamma$--ray connection in {\it Fermi} blazars.
{\it Mon. Not. R. Astr. Soc.} {\bf 413}, 852--862 (2011)

\bibitem{nemmen12} Nemmen, R.S., Georganopoulos, M., Guiriec, S., Meyer, E.T., Gehrels, N. \& Sambruna, R.M. 
A Universal Scaling for the Energetics of Relativistic Jets from Black Hole Systems.
{\it Science} {\bf 338,} 1445--1448  (2012)

\bibitem{mad} Tchekhovskoy, A., Metzger, B.D., Giannios, D. \& Kelley, L.Z. 
Swift J1644+57 gone MAD: the case for dynamically important magnetic flux 
threading the black hole in a jetted tidal disruption event.
{\it Mon. Not. R. Astr. Soc.} {\bf 437}, 2744--2760 (2014)

\bibitem{peterson00} Peterson, B.M., Wandel, A.  
Evidence for Supermassive Black Holes in Active Galactic Nuclei from Emission-Line Reverberation.
{\it Astrophys. J.} {\bf 540} L13--L16  (2000) 

\bibitem{mclure04} McLure, R.J. \& Dunlop, J.S. 
The cosmological evolution of quasar black hole masses.
{\it Mon. Not. R. Astr. Soc.} {\bf 352}, 1390--1404 (2004)

\bibitem{vestergaard06} Vestergaard, M. \& Peterson, B.M.  
Determining Central Black Hole Masses in Distant Active Galaxies and Quasars. II. 
Improved Optical and UV Scaling Relationships.
{\it Astrophys. J.} {\bf 641}, 689--709 (2006)

\bibitem{shakura73}	Shakura, N.I. \& Sunyaev, R.A. 
Black holes in binary systems. Observational appearance.
{\it Astron. Astrophys.} {\bf 24}, 337--355 (1973)

\bibitem{livio99}	Livio, M., Ogilvie, G.I. \& Pringle, J.E.
Extracting Energy from Black Holes: The Relative Importance of the Blandford-Znajek Mechanism.
{\it Astrophys. J.} {\bf 512}, 100--104 (1999)

\bibitem{meier02} Meier, D.L.
Grand Unification of AGN and the accretion and spin paradigms.
{\it New Astron. Rev.} {\bf 46}, 247--255 (2002)

\bibitem{tchekhovskoy12} 	Tchekhovskoy, A., McKinney, J.C. \& Narayan, R.,
    General Relativistic Modeling of Magnetized Jets from Accreting Black Holes.
    {\it J. of Physics: Conference Series} {\bf 372}, Issue 1, id. 012040 (2012).
	
\bibitem{sikora07} Sikora, M., Stawarz, L. \& Lasota, J.--P.
        Radio Loudness of Active Galactic Nuclei: Observational Facts and Theoretical Implications.
       {\it Astrophys. J.} {\bf 658}, 815--828 (2007)

\end{thebibliography}

\begin{thebibliography}{30}  
\setcounter{enumiv}{30} 
\bibitem{abdo10} Abdo, A.A., Ackermann, M., Ajello, M. et al. 
        The First Catalog of Active Galactic Nuclei Detected by the Fermi Large Area Telescope.
        {\it Astrophys. J.} {\bf 715}, 429--457 (2010) 

\bibitem{shen11} Shen, Y., Richards, G.T., Strauss, M.A. et al.
         A Catalog of Quasar Properties from Sloan Digital Sky Survey Data Release 7.
         {\it Astrophys. J. Supp.} {\bf 194}, 45 (21 pp) (2011)

\bibitem{ackermann11} Ackermann, M., Ajello, M., Allafort, A. et al.
	    The Second Catalog of Active Galactic Nuclei Detected by the Fermi Large Area Telescope.
        {\it Astrophys. J.} {\bf 743}, 171 (37 pp) (2011) 

\bibitem{massefit1} Ghisellini, G.
         Extragalactic relativistic jets.
         25th Texas Symp.. AIP Conf. Proceed. {\bf 1381}, 180--198 (2011) (arXiv:1104.0006) 

\bibitem{massefit2} Sbarrato, T., Ghisellini, G., Nardini, M., Tagliaferri, G., Greiner, J., Rau, A. \& Schady, P.
         Blazar candidates beyond redshift 4 observed with GROND
         {\it Mon. Not. R. Astr. Soc.} {\bf 433}, 2182--2193 (2013)


\bibitem{fossati98} Fossati, G., Maraschi, L., Celotti, A., Comastri, A. \& Ghisellini, G. 
         A unifying view of the spectral energy distributions of blazars.
         {\it Mon. Not. R. Astr. Soc.}  {\bf 299}, 433--448 (1998)

\bibitem{donato01} Donato, D., Ghisellini, G., Tagliaferri, G. \& Fossati, G. 
	Hard X--ray properties of blazars.
	{\it Astron. Astrophys.} {\bf 375}, 739--751 (2001)

\bibitem{bonnoli11} Bonnoli, G., Ghisellini, G., Foschini, L, Tavecchio, F. \& Ghirlanda, G.
         The $\gamma$--ray brightest days of the blazar 3C 454.3.
          {\it Mon. Not. R. Astr. Soc.} {\bf 410}, 368--380 (2011)

\bibitem{ghisellini09} Ghisellini, G. \& Tavecchio, F. 
	     Canonical high--power blazars.
        {\it Mon. Not. R. Astr. Soc.} {\bf 397}, 985--1002 (2009) 

\bibitem{nalewajko14} Nalewajko, K., Begelman, M.C. \& Sikora, M. 
        Constraining the Location of Gamma-Ray Flares in Luminous Blazars.
        {\it Astrophys. J.} {\bf 789}, 161, 20pp  (2014)

\bibitem{bottcher13} B\"ottcher, M., Reimer, A., Sweeney, K. \& Prakash, A.
          Leptonic and Hadronic Modeling of Fermi-detected Blazars.
          {\it Astrophys. J.} {\bf 768}, 54  14 pp  (2013)   
          
\bibitem{bentz06}  Bentz, M.C., Peterson, B.M., Pogge, R.W., Vestergaard, M. \& Onken, C.A.
                  The Radius--Luminosity Relationship for Active Galactic Nuclei: 
                  The Effect of Host--Galaxy Starlight on Luminosity Measurements.
                  {\it Astroph. J.} {\bf 644}, 133--142 (2006)
                  
\bibitem{koshida14} Koshida, S., Minezaki, T., Yoshii, Y., et al. 
     Reverberation measurements of the inner radius of the dust torus in 17 Seyfert galaxies.
     {\it Astrophys. J.} {\bf 788}, 159, 21pp  (2014)  

\bibitem{dermer95} Dermer, C.  
        On the Beaming Statistics of Gamma-Ray Sources.
        {\it Astrophys. J.} {\bf 446}, L63--L66 (1995)
        
\bibitem{sikora00} Sikora, M. \& Madejski, G.
     On Pair Content and Variability of Subparsec Jets in Quasars.
        {\it Astrophys. J.} {\bf 534}, 109--113 (2000)
   
\bibitem{tavecchio98} Tavecchio, F., Maraschi, L. \& Ghisellini, G.
        Constraints on the Physical Parameters of TeV Blazars.
       {\it Astrophys. J.} {\bf 509}, 608--619 (1998)

\bibitem{ghisellini01} Ghisellini, G. \&  Celotti, A.  
       Relativistic large--scale jets and minimum power requirements.
       {\it Mon. Not. R. Astr. Soc.} {\bf 327}, 739--743 (2001)

\bibitem{ghisellini96} Ghisellini, G. \&  Madau, P.
         On the origin of the gamma-ray emission in blazars.
         {\it Mon. Not. R. Astr. Soc.} {\bf 280}, 67--76 (1996)

\bibitem{ghisellini12}  Ghisellini G.  
         Electron-positron pairs in blazar jets and $\gamma$--ray loud radio galaxies.
          {\it Mon. Not. R. Astr. Soc.} {\bf 424}, L26--L30 (2012)

\end{thebibliography}
\end{document}